\documentclass[11pt,a4paper]{article}
\pdfoutput=1
\usepackage{jcappub}
\usepackage[T1]{fontenc}
\usepackage{color}
\usepackage{graphicx}
\usepackage{bm}
\usepackage[utf8]{inputenc}
\usepackage{comment}
\usepackage{amsmath,amsfonts,amssymb}
\usepackage{braket}
\usepackage{subcaption}
\usepackage{hyperref}
\usepackage[normalem]{ulem}

\title{Big Bang Nucleosynthesis constraints on sterile neutrino and lepton asymmetry of the Universe}

\author[a]{Graciela B. Gelmini,}
\author[b,c]{Masahiro Kawasaki,}
\author[a,c]{Alexander Kusenko,}
\author[b,c]{Kai Murai,}
\author[a]{Volodymyr Takhistov}
\affiliation[a]{Department of Physics and Astronomy, University of California, Los Angeles, CA 90095-1547, USA}
\affiliation[b]{Institute for Cosmic Ray Research, The University of Tokyo, Kashiwa 277-8582, Japan}
\affiliation[c]{Kavli Institute for the Physics and Mathematics of the Universe (WPI), The University of Tokyo, Kashiwa 277-8583, Japan}

\abstract{We consider the cosmological effects of sterile neutrinos with the masses of $150- 450$~MeV.
The decay of sterile neutrinos changes the thermal history of the Universe and affects the energy density of radiation at recombination and the big bang nucleosynthesis (BBN) results. We derive severe constraints on the parameters of sterile neutrinos from the primordial abundances of helium-4 and deuterium. We also find that in a particular model the constraints can be considerably relaxed by assuming a large lepton asymmetry in the active neutrinos.
In this case, the consistent parameters result in $N_{\mathrm{eff}} \simeq 3.2- 3.4$ and can alleviate the Hubble tension.}


\emailAdd{kawasaki@icrr.u-tokyo.ac.jp}
\emailAdd{kmurai@icrr.u-tokyo.ac.jp}

\begin{document}
\baselineskip 0.58cm
\begin{flushright}
    IPMU 20-0051
\end{flushright}
 \maketitle

\section{Introduction}
\label{sec:intro}
The discovery of neutrino oscillations~\cite{Fukuda:1998mi} has firmly established the existence of neutrino masses and flavor mixing. As weakly interacting (active) neutrinos are massless within the Standard Model (SM), this motivates new  physics in the neutrino sector. While precision measurements of Z-boson decays restrict the number of active neutrino species to three~\cite{ALEPH:2005ab}, this  does not exclude the presence of additional sterile neutrinos that are singlets under the SM gauge groups.

Sterile neutrinos have been extensively discussed in the literature, as  they are motivated by various theoretical considerations and play a central role in models for the origin of the neutrino masses. In particular, they are essential for the seesaw mechanism~\cite{Minkowski:1977sc,Yanagida:1979as,Mohapatra:1979ia,GellMann:1980vs}, which utilizes sterile neutrinos much heavier than the electroweak scale to generate small masses for active neutrinos. While the three-flavor neutrino paradigm has been extensively tested, a slew of anomalous results consistent with the existence of additional sterile neutrino species of $m_s \sim \mathcal{O}(\text{eV})$ mass have been reported, including from MiniBooNE~\cite{Aguilar-Arevalo:2018gpe} and LSND~\cite{Aguilar:2001ty} short baseline experiments, gallium experiments~\cite{Giunti:2010zu,Kostensalo:2019vmv}, measurements of reactor neutrino flux~\cite{Mention:2011rk} as well as combined fits~\cite{Dentler:2018sju,Gariazzo:2018mwd,Liao:2018mbg} to data from NEOS~\cite{Ko:2016owz} and DANSS~\cite{Alekseev:2018efk} reactor experiments.
Sterile neutrinos with mass of $m_s \sim \mathcal{O}(\text{keV})$ have been investigated in the context of dark matter
~\cite{Dodelson:1993je,Shi:1998km,Dolgov:2000ew,Abazajian:2001nj,Abazajian:2001vt,Asaka:2005an,Asaka:2005pn,Shaposhnikov:2006xi,Kusenko:2006rh,Petraki:2007gq,Petraki:2008ef,Loewenstein:2008yi,Loewenstein:2009cm,Kusenko:2010ik,Loewenstein:2012px,Adhikari:2016bei,Abazajian:2019ejt}. Furthermore, keV sterile neutrinos can also play a role in supernova explosions~\cite{Kusenko:1997sp,Kusenko:1998bk,Fuller:2009zz} and in generating the matter-antimatter asymmetry of the universe~\cite{Fukugita:1986hr,Luty:1992un,Akhmedov:1998qx,Asaka:2005pn}. For an overview of sterile neutrinos see, e.g.,  Ref.~\cite{Kusenko:2009up}.

While sterile neutrinos with sub-MeV mass have already garnished significant consideration, heavier sterile neutrinos have not been as well explored.
Heavy sterile neutrinos with mass $m_s \lesssim$ GeV that mix with active neutrinos with strength below the current laboratory bounds and decaying around the time of Big Bang Nucleosynthesis (BBN) can affect the thermal history of the Universe in a profound manner.
Previously, in~Ref.~\cite{Fuller:2011qy}, one of the present authors and collaborators derived the constraints on heavy sterile neutrino mass and mixing parameters from the Cosmic Microwave Background (CMB) limit on the effective number of relativistic neutrino species $N_{\mathrm{eff}}$ (see Ref.~\cite{Hansen:2003yj} for another possible effect on the CMB of sterile neutrinos in a similar mass range).
However, it is also paramount to address constraints from BBN, as sterile neutrinos decaying around the BBN epoch can dramatically affect the abundances of the synthesized light elements (D and $^4$He).
The lifetime of heavy sterile neutrinos has been roughly restricted from above by BBN considerations~\cite{Boyarsky:2009ix} (see also~Ref.~\cite{Bolton:2019pcu}), but detailed analyses of BBN effects and constraints were only performed for sterile neutrinos lighter than the pion~\cite{Dolgov:2000pj,Dolgov:2000jw,Ruchayskiy:2012si,Vincent:2014rja}.

In this paper, we investigate the effect on BBN of heavy sterile neutrinos with the mass $150~\mathrm{MeV} < m_s < 450~\mathrm{MeV}$ that mix only with electron neutrinos and decay just after $e^+ e^-$ annihilation.
Lighter sterile neutrinos, with $m_s < 150$ MeV, are severely constrained by the CMB limit on $N_{\mathrm{eff}}$~\cite{Fuller:2011qy}.
For heavier ones, with $m_s > 450$ MeV, one needs to include additional decay channels of sterile neutrinos into kaons, eta, or 3 pions.

We find that BBN imposes stringent constraints on the allowed sterile neutrino parameter space.
Further, we point out that the BBN limits can be relaxed if there is a large lepton asymmetry before the BBN epoch, and study this situation in a particular model.

This paper is organized as follows.
In Sec.~\ref{sec:sterile in the early universe}, we summarize the behavior of the sterile neutrinos in the early Universe. In Sec.~\ref{sec:sterile decay}, we review the decay of sterile neutrinos and its effects on cosmology.
We give the results of the BBN calculations with the sterile neutrino in Sec.~\ref{sec:sterile BBN w/o lepton asymmetry} and the results of a particular model with lepton asymmetries in Sec.~\ref{sec:sterile BBN w/ lepton asymmetry}.
Finally, our conclusions and discussions are in Sec.~\ref{sec:result}.

\section{Sterile neutrinos in the early Universe}
\label{sec:sterile in the early universe}
The existence of the sterile neutrinos changes the cosmological scenario.
In the following, we consider how the sterile neutrinos affect the thermal history of the Universe.

\subsection{Production and decoupling}
We assume that the sterile neutrinos are produced exclusively through their  mixing with active neutrinos\footnote{We note that sterile neutrinos could also be produced through other mechanisms beyond active-sterile neutrino mixing, such as decays of additional heavy scalars in non-minimal particle models~\cite{Kusenko:2006rh,Petraki:2007gq}.} while they are relativistic, i.e. when the cosmic temperature is
\begin{equation}
  T > m_s~.
\end{equation}
First, we will check if sterile neutrinos were  in thermal equilibrium at high temperatures. In this case, the production rate of the sterile neutrinos through their mixing with the electron neutrino should be larger than the cosmic expansion rate, the Hubble parameter $H$. Since the sterile neutrinos are produced through the weak interactions and the mixing,
the production rate $\Gamma_{\mathrm{prod}}$ can be estimated by $\Gamma_{\mathrm{prod}} \simeq G_F^2T^5\sin ^2 \theta_{\mathrm{M}} $, where $G_F$ is the Fermi constant and $\theta_{\mathrm{M}}$ is the active-sterile neutrino mixing angle in the medium. The mixing angle in the medium is related to the mixing angle in the vacuum $\theta$ as
\begin{equation}
  \sin^2 \theta_{\mathrm{M}} \sim \frac{\sin^2 \theta}{\left[1 + 9.6 \times 10^{-24}(T/\mathrm{MeV})^6 (m_s/150\mathrm{MeV})^{-2}\right]^2}~,
  \label{eq:matter effect}
\end{equation}
in the limit of small mixing~\cite{Gelmini:2019wfp}.
Thus, matter effects become important when
\begin{equation}
  T \gtrsim 6.9~\mathrm{GeV} \left(\dfrac{m_s}{150\mathrm{MeV}}\right)^{1/3}~.
  \label{eq:matter effect condition}
\end{equation}

Therefore, the condition for sterile neutrinos to be in thermal equilibrium is
\begin{equation}
\label{eq:equilibrium}
  \frac{\Gamma_{\mathrm{prod}}}{H} \simeq
  \frac{G_F^2T^5\sin ^2 \theta_{\mathrm{M}}}{\sqrt{\dfrac{\pi^2 g_*}{90}}\dfrac{T^2}{M_{\rm Pl}}}>1~.
\end{equation}
Here $M_{\rm Pl}\simeq 2.4\times 10^{18}$~GeV is the (reduced) Planck mass.
By rewriting this as a condition for the temperature, we obtain
\begin{equation}
  T \gtrsim \left(\frac{\pi^2 g_*}{90}\right)^{1/6}G_F^{-2/3}M_{\mathrm{Pl}}^{-1/3}\
  \left( \frac{1}{\sin ^2 \theta_{\mathrm{M}}} \right)^{1/3}
  \simeq 4.5~\mathrm{GeV}
  \left(\frac{g_*}{86.25} \right)^{1/6}
  \left(\frac{10^{-10}}{\sin ^2 \theta_{\mathrm{M}}}\right)^{1/3} = T_{\mathrm{f.o.}}~,
  \label{eq:sterile thermal condition}
\end{equation}
where we take $g_* = 86.25$ as a reference value for the number of degrees of freedom in the thermal bath that corresponds to $m_b < T < m_W$, where
$m_b = 4.2~\mathrm{GeV}$ is the mass of the bottom quark and $m_W = 80.4~\mathrm{GeV}$ is the mass of the W gauge boson. If the condition in Eq.~\eqref{eq:equilibrium} is satisfied at high temperatures $T> T_\mathrm{f.o.}$, then $T_\mathrm{f.o.}$ is the decoupling or freeze-out temperature.
As we can see from Eq.~\eqref{eq:sterile thermal condition}, sterile neutrinos are already decoupled at $T \sim m_b$ for $\sin^2 \theta < 10^{-10}$, because $\sin^2 \theta_{\mathrm{M}} < \sin^2 \theta$.
Therefore, the value $g_* = 86.25$ is justified for $\sin^2 \theta < 10^{-10}$.
By comparing Eqs.~(\ref{eq:matter effect condition}) and (\ref{eq:sterile thermal condition}), we find that sterile neutrinos are thermally produced and the matter effects are negligible if $\sin^2 \theta \gtrsim 2.8 \times 10^{-11} (m_s/150~\mathrm{MeV})^{-1}$.

On the other hand, for a smaller mixing, matter effects become important and sterile neutrinos may not be thermally produced. In this case, sterile neutrinos are non-thermally produced through active-sterile neutrino oscillations, which is known as the Dodelson-Widrow (DW) mechanism~\cite{Dodelson:1993je}.
The produced sterile neutrinos are less abundant than those produced thermally and experience entropy dilution after their production, which happens mostly at a temperature $T_{\mathrm{max}}$ at which the DW production rate is maximum. However, these sterile neutrinos can still affect the BBN and thus can be constrained by BBN.
Considering the entropy dilution after $T_{\mathrm{max}}$, the number density at $T\sim 1$ MeV of non-thermally produced sterile neutrinos at $T\sim 1$ MeV can be written as~\cite{Gelmini:2019wfp,Rehagen:2014vna}
\begin{equation}
	n_{\nu_s}(T)
	= \left(1 - \exp \left[-3.27~ \left(\frac{\sin^2 \theta}{10^{-10}}\right) \left( \frac{m_s}{150~\mathrm{MeV}} \right)\right] \right)\dfrac{g_{s*}(T)}{g_{s*,\mathrm{prod}}} n_{\nu_{\alpha}}(T)
    \label{eq:DW sterile number density}
\end{equation}
where $n_{\nu_{\alpha}}$ is the number density of one active neutrino species and $g_{s*,\mathrm{prod}} = 86.25$ is the number of relativistic degrees of freedom when sterile neutrinos are produced. The DW production rate is maximum at $T_{\mathrm{max}} \simeq 5.7~\mathrm{GeV} (m_s/150~\mathrm{MeV})^{1/3}$, when $\sin \theta_{\mathrm{M}}/\sin \theta = 1/1.34$~\cite{Gelmini:2019clw}. We can consider that sterile neutrinos are mainly produced at $T_{\mathrm{max}}$ and use $g_{s*, {\rm prod}} = 86.25$ for our purpose.
From the BBN upper bound of $N_{\mathrm{eff}}<3.4$~\cite{Cyburt:2015mya} (see also~Particle Data Group~\cite{Tanabashi:2018oca}), we can conclude that sterile neutrinos are constrained by BBN when $\rho_{\nu_s}=m_s n_{\nu_s} > 0.4 \rho_{\nu_{\alpha}}$ at BBN ($T \sim 1$~MeV).
Using Eq.~(\ref{eq:DW sterile number density}) when the exponent is small, this condition implies

\begin{equation}
    \sin ^2 \theta > 2.0 \times 10^{-12} \left(\frac{m_s}{150~\mathrm{MeV}}\right)^{-2}.
\end{equation}

\subsection{Energy density of the non-relativistic sterile neutrino}

We now calculate the energy density of the sterile neutrinos before they decay. Here, we focus on the sterile neutrinos that decay at temperatures lower than the MeV scale. As we show in detail in Sec.~\ref{sec:annihilation of decay product neutrino} and Appendix~\ref{sec:annihilation detail}, in this case the re-scattering of active neutrinos produced in the decay is negligible. Thus, sterile neutrinos with mass $m_s > 150$~MeV behave like matter when they decay.

The energy density of non-relativistic sterile neutrinos $\rho_{\nu_s}$ when they decay is
\begin{equation}
  \rho_{\nu_s}(T_{\mathrm{decay}}) = m_s~n_{\nu_s}(T_{\mathrm{decay}}),
  \label{eq:sterile neutrino energy density}
\end{equation}
where $T_{\mathrm{decay}}$ is the cosmic temperature when sterile neutrinos decay.
\begin{figure}[t]
  \begin{center}
    \includegraphics[clip,width=0.8\textwidth]{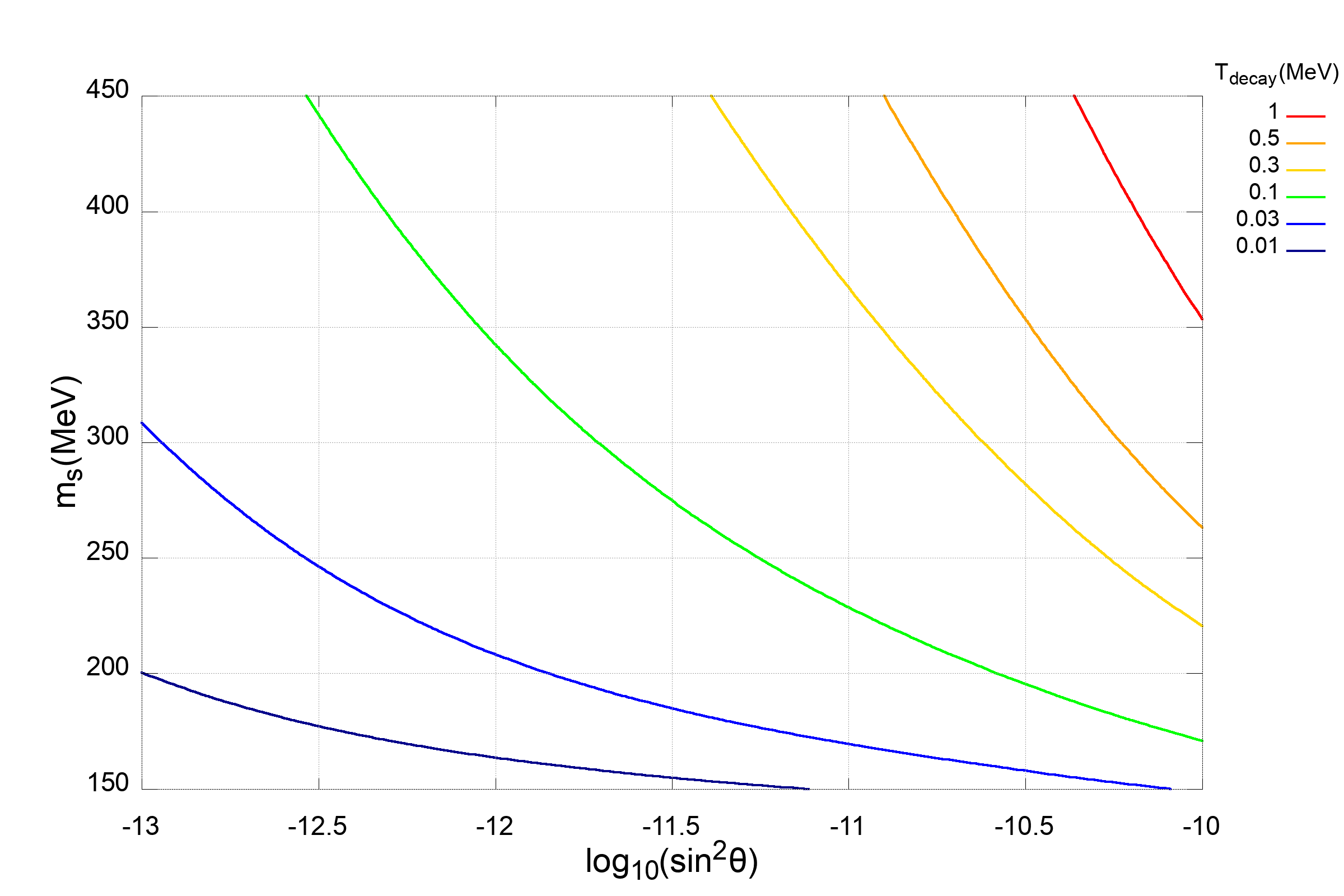}
    \caption{Temperature of the Universe when the sterile neutrinos decay.
    Note that, since we assume the decay occurs after the electron-positron annihilation we consider only the region in which $T_{\rm decay} < 0.5$~MeV.}
    \label{fig:decaytemperature}
  \end{center}
\end{figure}

The non-relativistic sterile neutrinos can dominate the Universe before they decay, when
\begin{equation} \label{eq:BBNlim}
  \Gamma_{\mathrm{decay}} = H(T),
\end{equation}
where the Hubble parameter is
\begin{equation}
  H(T) = \sqrt{\frac{\rho_{\nu_s}+\rho_R}{3M_{\mathrm{Pl}}^2}}~
  =\frac{1}{\sqrt{3}M_{\mathrm{Pl}}}
  \left[
    m_s n_{\nu_s}(T)
   +
    g_*(T) \frac{\pi^2}{30}T^4
  \right]^{\frac{1}{2}}~.
\end{equation}
Considering only sterile neutrinos that decay after the electron-positron annihilation, and thus $g_{s*}(T_{\mathrm{decay}}) = 43/11$ and $g_*(T_{\mathrm{decay}}) = 2+(21/4)(4/11)^{4/3}$, the decay temperature is shown in Fig.~\ref{fig:decaytemperature}.
In the following, we focus on the regions where $T_{\mathrm{decay}} < m_e\simeq 0.5$~MeV.

\begin{figure}[t]
  \begin{center}
    \includegraphics[clip,width=0.8\textwidth]{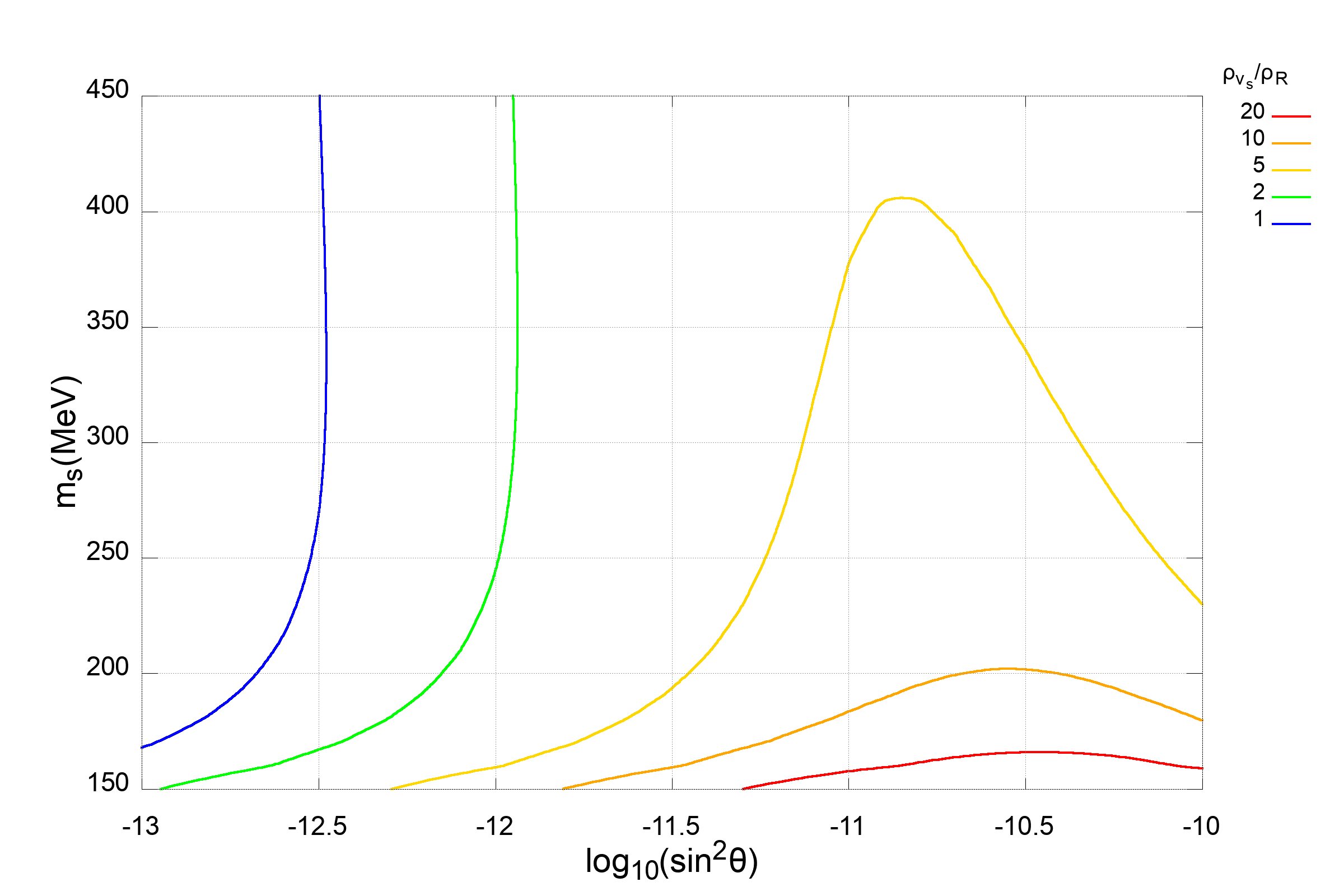}
    \caption{Ratio of sterile neutrino and radiation energy densities when the sterile neutrinos decay.
    }
    \label{fig:energyratio}
  \end{center}
\end{figure}

Using $T_{\rm decay}$, we calculate the energy ratio of sterile neutrinos and  radiation at decay. The results, plotted in Fig.~\ref{fig:energyratio}, show that the Universe could have been dominated by the non-relativistic sterile neutrinos just before they decayed.

\section{Sterile neutrino decay}
\label{sec:sterile decay}
As mentioned above, heavy sterile neutrinos decay after the decoupling of active neutrinos. Since the sterile neutrinos inject their energy into both the plasma and the neutrino sector through their decays, these cause entropy production and change $N_{\mathrm{eff}}$.

\begin{figure}[t]
  \hspace{1.2cm}(a)\hspace{8.5cm}(b)\\ \vspace{-1.34cm}
  \begin{center}
    \includegraphics[clip,width=0.55\textwidth]{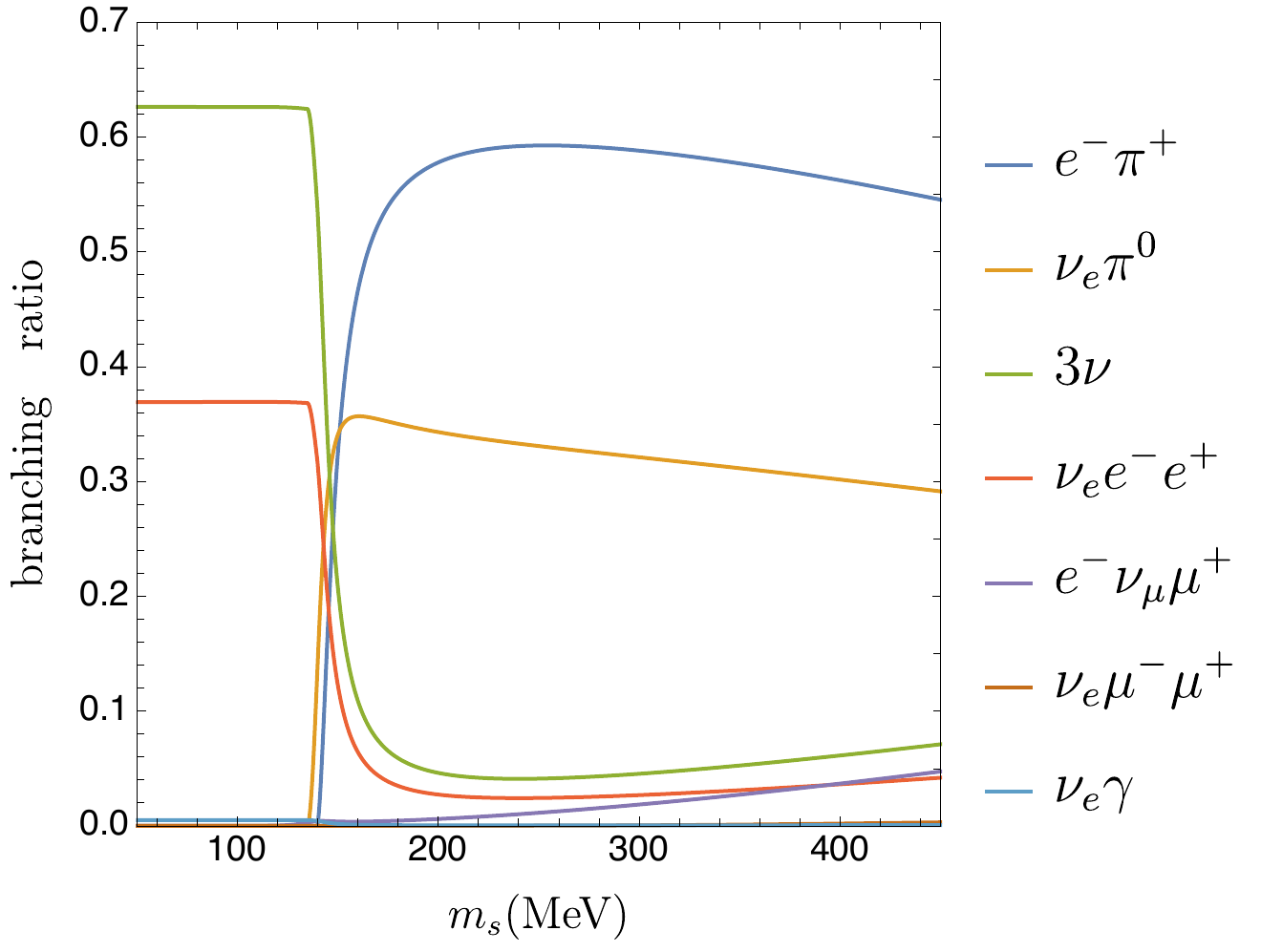}
    \includegraphics[clip,width=0.43\textwidth]{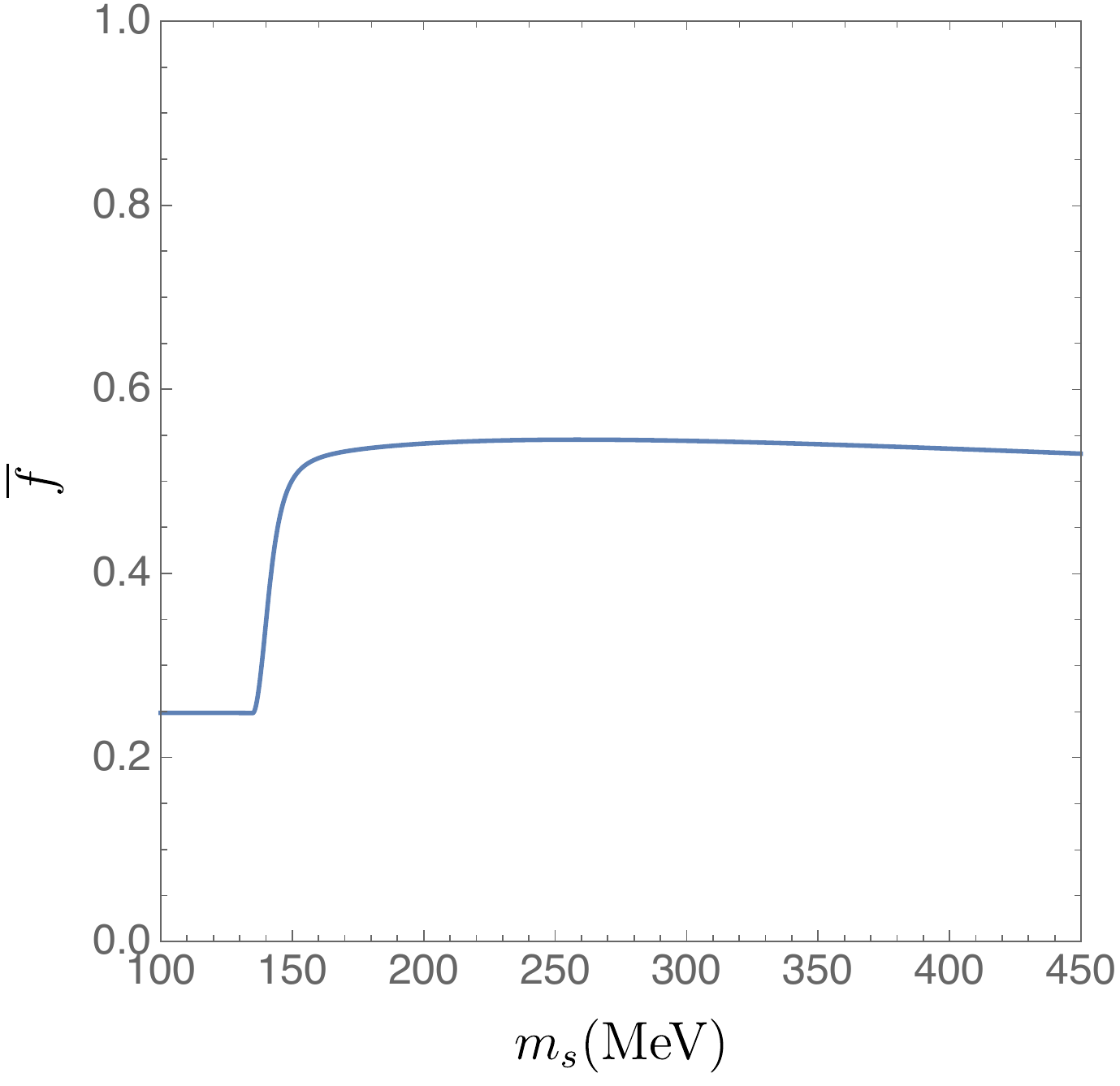}
    \caption{(a) Branching ratios of the sterile neutrino decay modes. They depend only on $m_s$ (and not on $\sin \theta$).~(b) The average fraction $\overline{f}$ of the sterile neutrino rest mass injected into the plasma.}
    \label{fig:branch_f}
  \end{center}
\end{figure}

The decay channels we consider are summarized in App.~\ref{sec:decay channel}.~For each channel, the decay rate $\Gamma$ and the energy fraction $f$ injected into the plasma are also shown in App.~\ref{sec:decay channel}.
Using them for each channel, the total sterile neutrino decay rate and the average $\overline{f}$ can be obtained.~The decay channels in Appendix~\ref{sec:decay channel} are all the processes where a sterile neutrino produces an electron neutrino through their mixing.~Therefore, the decay rates are all proportional to $\sin^2 \theta$ and the branching ratios depend only on $m_s$.~We show the branching ratios in Fig.~\ref{fig:branch_f} (a) and $\overline{f}$ in~Fig.~\ref{fig:branch_f} (b).
Here, we assume that the sterile neutrinos decay after active neutrino decoupling and the neutrinos produced in the decay do not inject energy into the plasma sector.
Although the neutrinos with high energy do inject some fraction of their energy into the plasma through annihilations or scatterings or modify the neutron-proton ratio through weak interaction~\cite{Dolgov:2000pj}, as we discuss in Sec.~\ref{sec:annihilation of decay product neutrino}, we will restrict ourselves to the regions in the mass and mixing parameter space where these effects are negligible.

\begin{figure}[t]
  \begin{center}
    \includegraphics[clip,width=0.48\textwidth]{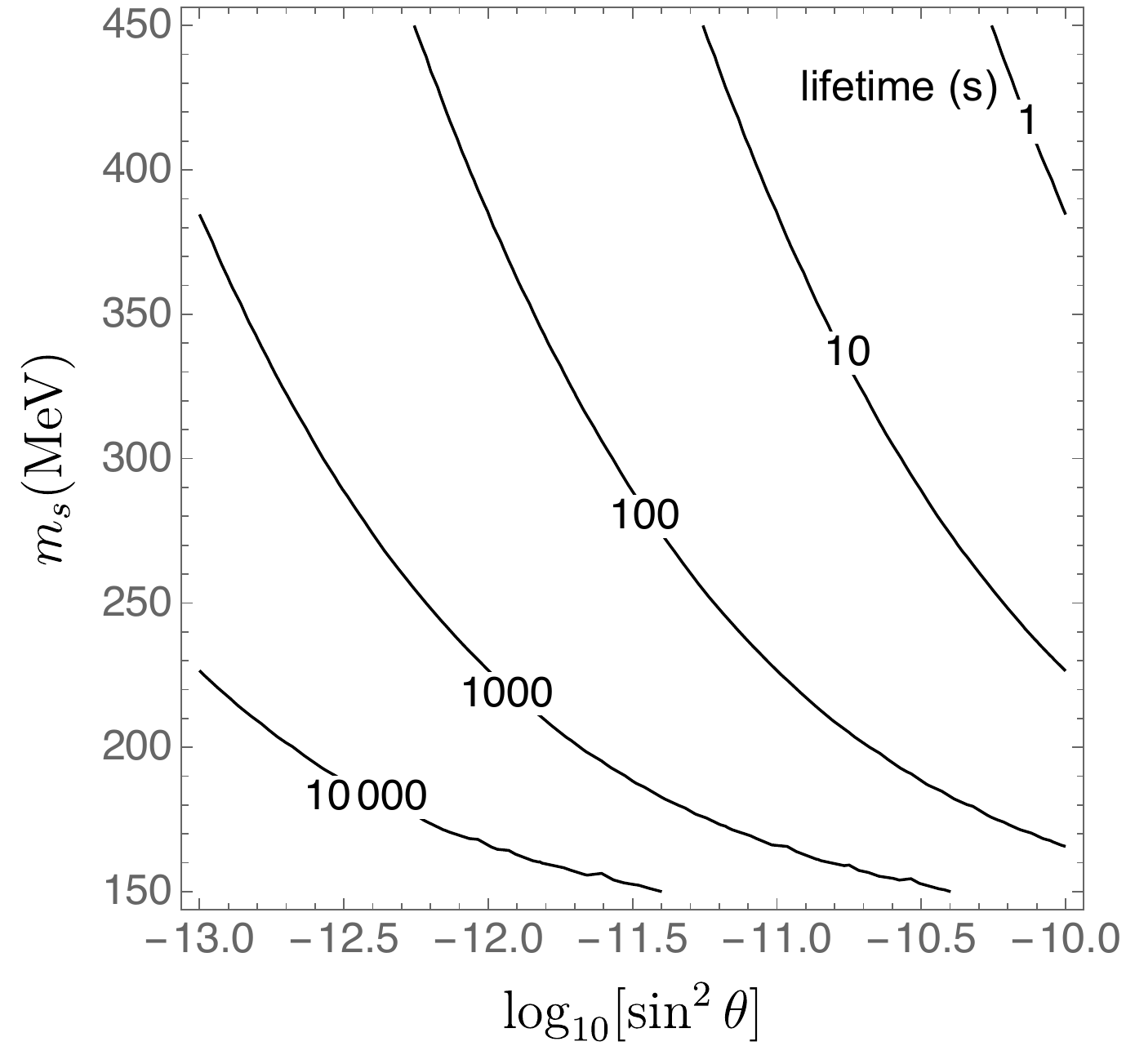}
    \caption{The sterile neutrino lifetime.}
    \label{fig:lifetime}
  \end{center}
\end{figure}

The sterile neutrino lifetime is shown in Fig.~\ref{fig:lifetime}.
As we can see from the decay rate for each channel in App.~\ref{sec:decay channel}, the heavier the sterile neutrino is and the larger the mixing angle is, the shorter is the lifetime.

We note that, in contrast to Ref.~\cite{Fuller:2011qy}, in our study we employ the more detailed calculations of Ref.~\cite{Gorbunov:2007ak} for the rates of the relevant sterile neutrino decay channels. The resulting entropy generation, dilution and change in effective number of relativistic neutrino species $N_{\rm eff}$ from our analysis are qualitatively in agreement with those of Ref.~\cite{Fuller:2011qy}.

\subsection{Scattering of the high-energy neutrinos from the decay}
\label{sec:annihilation of decay product neutrino}
\begin{figure}[t]
  \begin{center}
    \includegraphics[clip,width=0.48\textwidth]{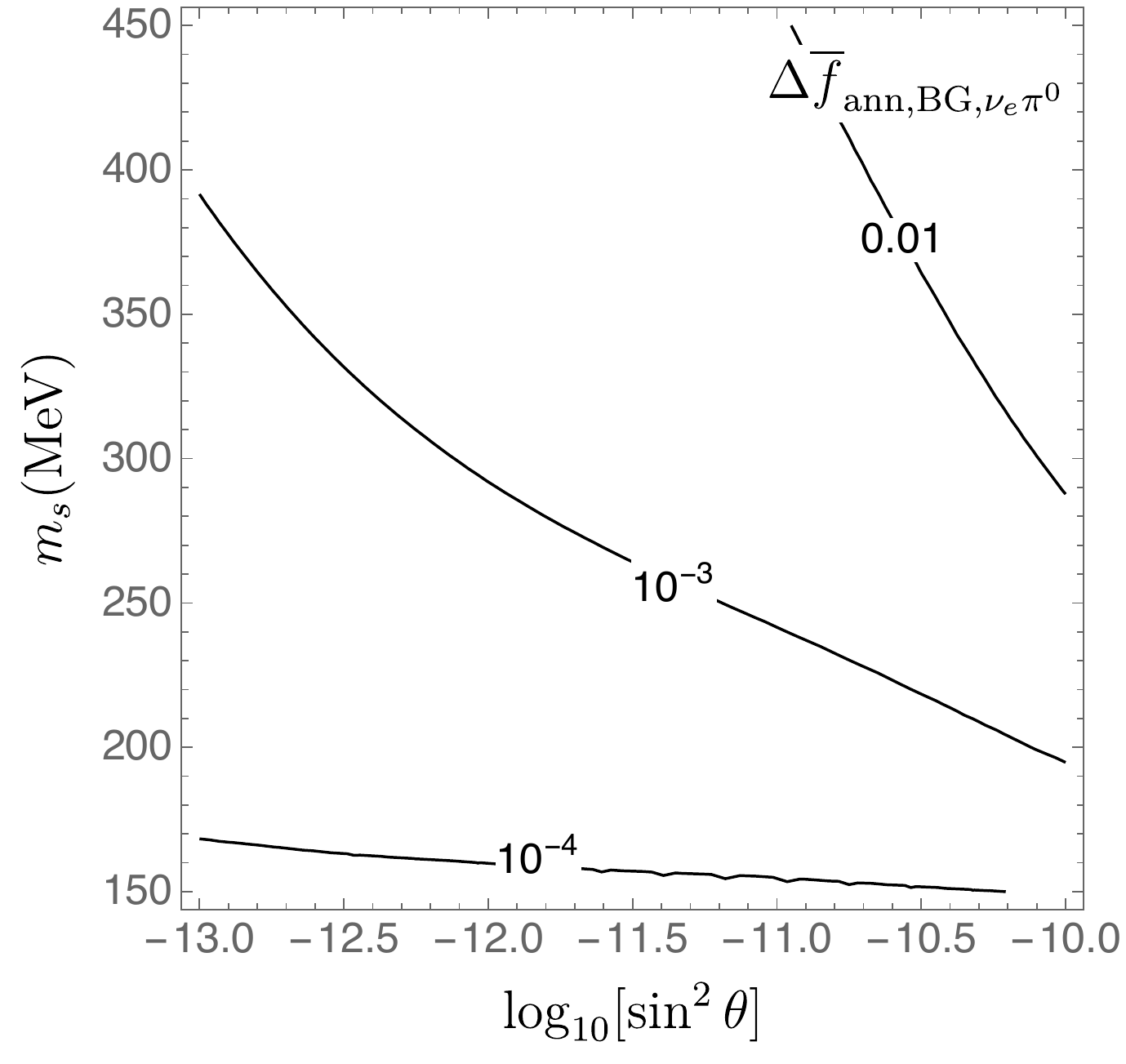}
    \includegraphics[clip,width=0.48\textwidth]{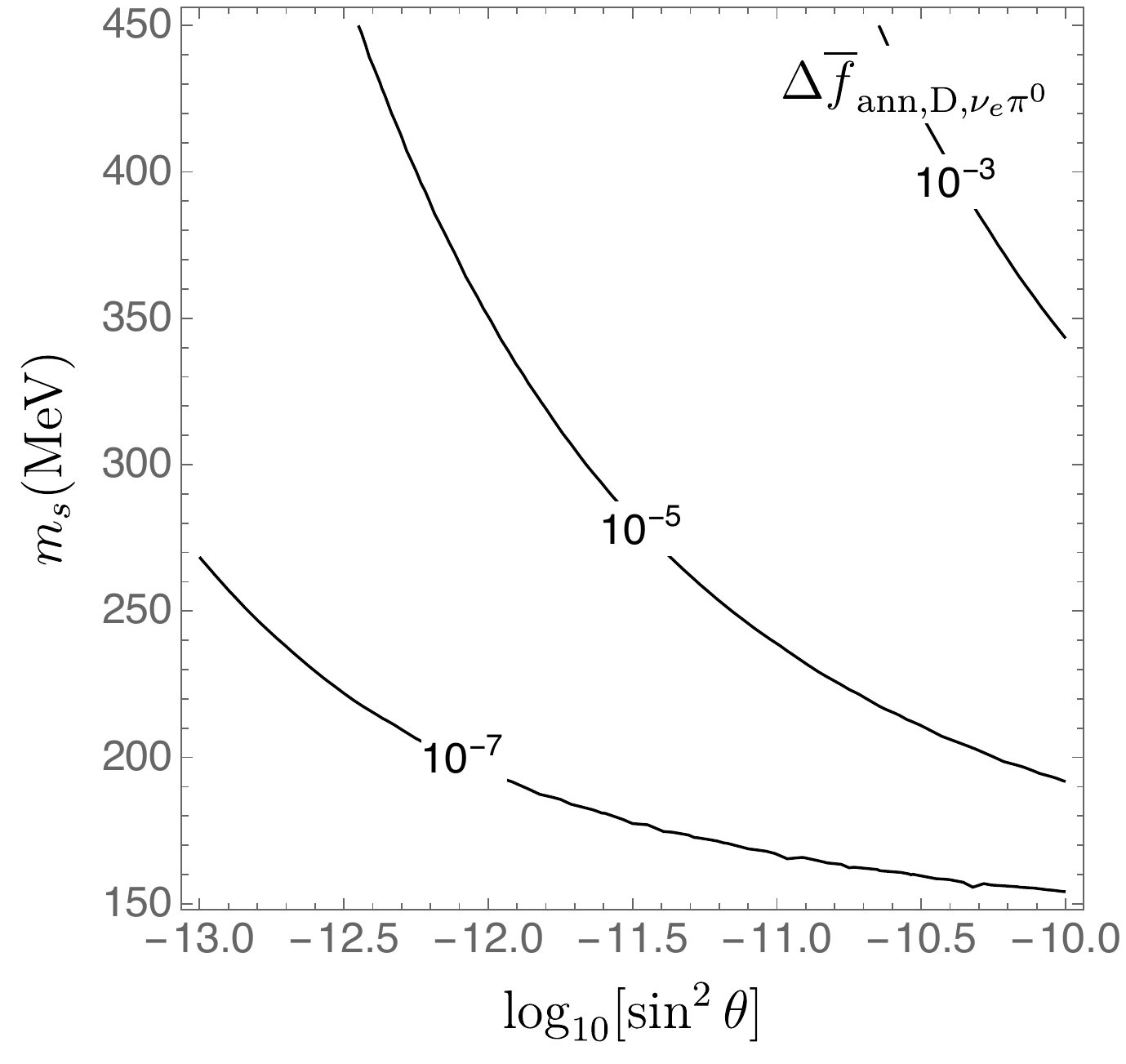}
    \caption{The correction to $\overline{f}$ due to the annihilation of the high-energy neutrinos produced from $\nu_s \to \nu_e + \pi^0$ process.
    The left panel shows the correction from the annihilation of the high-energy neutrinos and the background neutrinos.
    The right panel shows the correction from the annihilation of the high-energy neutrinos themselves.}
    \label{fig:f correction}
  \end{center}
\end{figure}

In the estimation of the average fraction $\overline{f}$ above, we assumed the neutrinos produced in the decay do not inject any energy into the thermal plasma sector.~Here we comment on a possible correction to $\overline{f}$ due to the interaction of the high energy neutrinos.
Since active neutrinos can produce electrons and positrons through weak interactions with other particles in the thermal bath, we can consider two situations:
\begin{itemize}
  \item The active neutrinos produced in the decay interact with the background particles.
  \item The active neutrinos produced in the decay interact with each other.
\end{itemize}
For each situation, we focus on the $\nu_s \to \nu_e + \pi^0$ decay process that is the dominant decay mode and estimate the correction to $\overline{f}$ due to the annihilation process $\nu_e + \overline{\nu}_e \to e^- + e^+$, shown in Fig.\ref{fig:f correction}. From the figure it is seen that the correction is smaller than $0.01$ in most of the parameter space and it is larger for scatterings off the background neutrinos.
In the following we only consider the parameter region satisfying $\Delta \overline{f}_{\mathrm{ann,BG,\nu_e\pi^0}} < 0.01$ in the left panel of Fig.\ref{fig:f correction} and hence neglect the correction. For a detailed discussion see App.~\ref{sec:annihilation detail}.

For the parameter region where $\bar{f}$ receives negligible corrections, the weak interaction processes involving high-energy neutrinos are also negligible and the modification of the neutron-proton ratio can be ignored.

\subsection{Entropy generation and dilution effect}
\label{sec:entropy production}

\begin{figure}[t]
  \begin{center}
    \includegraphics[clip,width=0.8\textwidth]{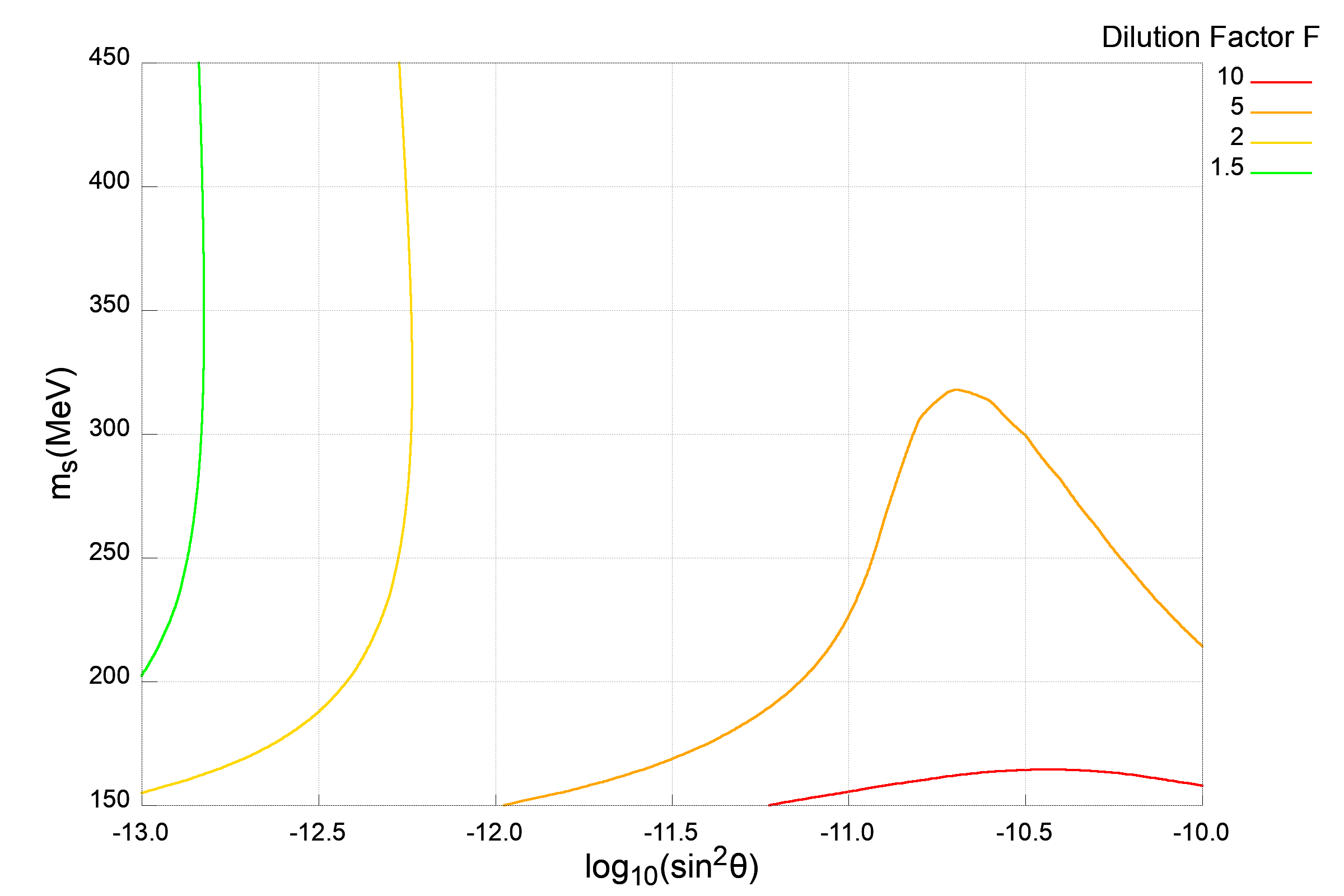}
    \caption{Ratio $F$ of the entropies after and before the sterile neutrino decay.
    }
    \label{fig:DilutionFactor}
  \end{center}
\end{figure}

The sterile neutrinos inject their energy into the plasma sector and the neutrino sector through their decay and generate entropy. Here we define the  ratio $F$ of the entropy after $(s_f)$ and before $(s_i)$ the decay
\begin{equation}
  F \equiv \frac{s_{f}}{s_{i}} = \frac{g_{f}T_f^3a_f^3}{g_{i}T_i^3a_i^3}~.
\end{equation}
The subscript $i$ indicates the time before the sterile neutrinos begin to decay and the subscript $f$ indicates the time after most of the sterile neutrinos decayed. We take the number of degrees of freedom of the plasma to be  $g_i=11/2$ and $g_f=2$. In other words, the decays occur after active neutrino decoupling and finish after the electron-positron annihilation.
Using $F$, we can compute the ratio of the photon and neutrino temperatures after the decay. Since $T_{\nu}$ satisfies $T_{\nu f}a_f = T_{\nu i} a_i$,
\begin{equation}
  \frac{T_{\nu f}}{T_f} =\frac{a_i}{a_f}\cdot\frac{T_i}{T_f}
  =\frac{1}{F^{1/3}}\left(\frac{g_{f}}{g_{i}}\right)^{1/3}
  =\frac{1}{F^{1/3}}\left(\frac{4}{11}\right)^{1/3},
\end{equation}
where we used $T_{\nu i} = T_i$.

This entropy generation has a significant impact on BBN. Considering it and the value of $\eta \equiv n_B/n_{\gamma}$ inferred from CMB observations, we can conclude that $\eta$ has a higher value in the BBN epoch than in the standard scenario.
We show entropy ratio $F$ values in the $(\sin^2 \theta, m_s)$ plane in Fig.~\ref{fig:DilutionFactor}.

\subsection{Change of \texorpdfstring{$N_{\mathrm{eff}}$}{}}
The decay of the sterile neutrino changes the effective number of neutrino species $N_{\mathrm{eff}}$, which is defined through the relation
\begin{equation}
  \rho_R = \left(
  1+\frac{7}{8}\left( \frac{4}{11} \right)^{4/3}N_{\mathrm{eff}}
  \right)\rho_{\gamma},
  \label{eq:definition of Neff}
\end{equation}
where $\rho_R$ is the total radiation energy density including the background and neutrinos produced due to the decays, and $\rho_\gamma$ is the photon energy density.
In our scenario, the decay of the sterile neutrinos injects some energy into the plasma and the neutrino sector.
Therefore, $N_{\mathrm{eff}}$ after the decay largely depends on the energy injection ratio $\overline{f}$.

\begin{figure}[t]
  \begin{center}
    \includegraphics[clip,width=0.8\textwidth]{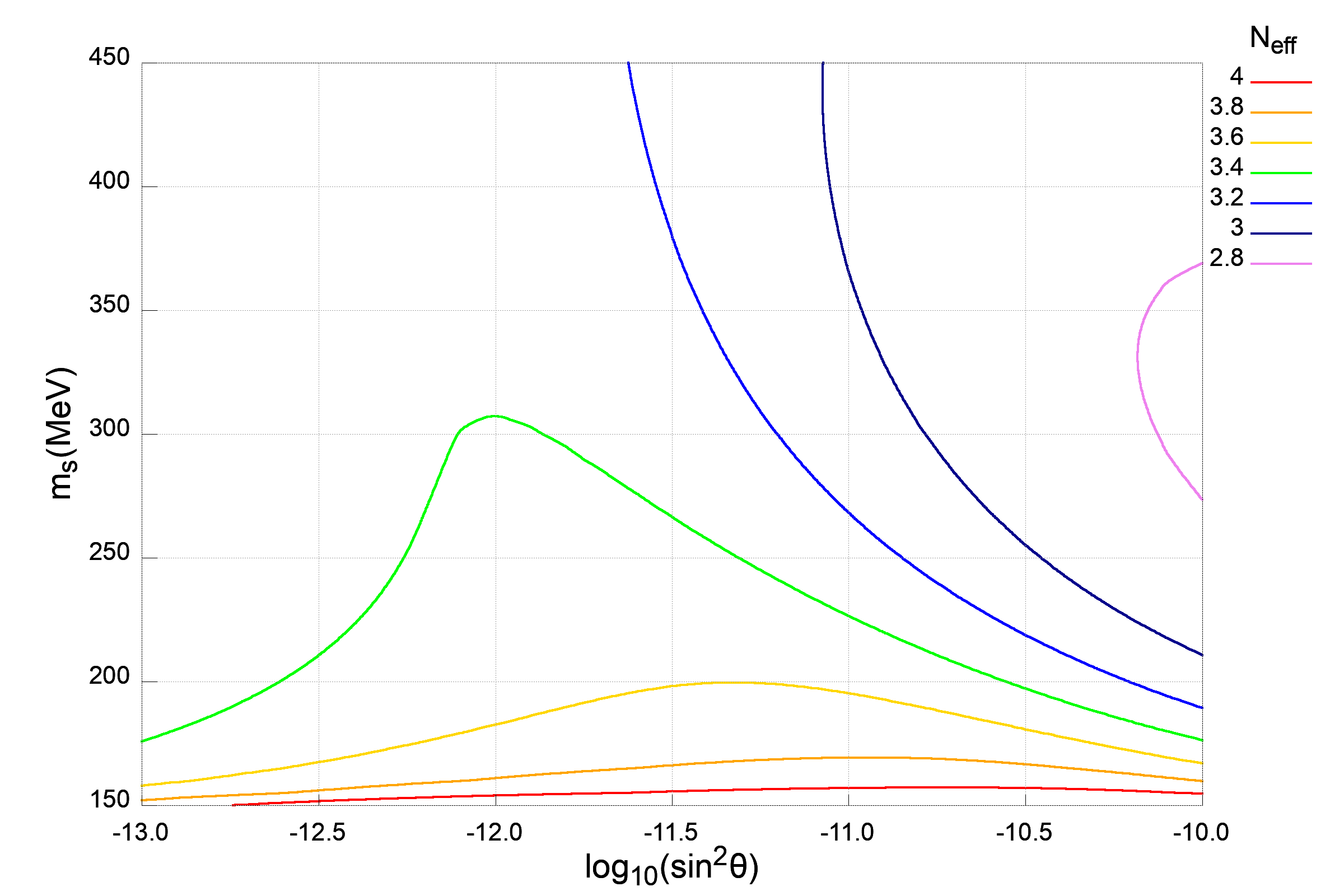}
    \caption{$N_{\mathrm{eff}}$ after the sterile neutrino decay.
    }
    \label{fig:Neff}
  \end{center}
\end{figure}

In Fig.~\ref{fig:Neff} we show $N_{\mathrm{eff}}$ after the decay of the sterile neutrinos. Our results confirm the observation of~Ref.~\cite{Fuller:2011qy} that $N_{\mathrm{eff}}$ can take values close to 3 in some region of the parameter space.
\section{BBN and constraint on the sterile neutrino}
\label{sec:sterile BBN w/o lepton asymmetry}
In this section, we show that the sterile neutrino parameters are tightly constrained by BBN considerations.
Following the usual BBN conventions, we write the abundances of the primordial $^4$He and D as $Y_p$ and D/H.
$Y_p$ is defined as the ratio of the $^4$He mass density and the total baryon density ($=\rho_{^4\text{He}}/\rho_B$) while D/H is the ratio of the D and H number densities~($=n_\text{D}/n_\text{H})$.

\subsection{Qualitative impact of the sterile neutrino}
First, we discuss qualitatively the effects of the sterile neutrinos on BBN.
The sterile neutrinos have a significant energy density before they decay, which increases the cosmic expansion faster.
The faster cosmic expansion during BBN results in an earlier freeze-out of the weak interactions and a larger neutron-to-proton ratio. This speed-up effect makes D/H and $Y_P$ larger.

As mentioned in Sec.~\ref{sec:entropy production}, the entropy generation by the sterile neutrino decay means a larger $\eta$ during the BBN epoch.
This effect makes D/H smaller and $Y_P$ larger.
However, if the sterile neutrinos decay before the synthesis of D ($T \simeq 0.1$~MeV),
the entropy generation does not significantly affect BBN.

\subsection{Observational constraints on the light elements}
Here we summarize observational constraints on the primordial abundances of D and $^4$He.
The deuterium abundance has been precisely determined by observing absorption spectra of QSOs due to damped Lyman-$\alpha$ systems.
Recently,  Cooke \textit{et al.}~\cite{Cooke:2017cwo} reported
\begin{equation}
    (\text{D}/\text{H})_p = (2.527 \pm 0.030)\times 10^{-5},
    \label{eq:D_obs}
\end{equation}
from measurements of the $7$ damped Lyman-$\alpha$ systems.
This estimation is consistent with $(\text{D}/\text{H})_p = (2.545 \pm 0.025)\times 10^{-5}$ obtained by Zavaryzin \textit{et al.}~\cite{Zavarygin:2018dbk}.
The Particle Data Group~\cite{Tanabashi:2018oca} suggests $(\text{D}/\text{H})_p =(2.547 \pm 0.025)\times 10^{-5}$ which is almost the same as that given in~\cite{Zavarygin:2018dbk}.

The primordial abundance of $^4$He is determined by measurements of recombination lines from extra-galactic HII regions.
Izotov \textit{et al.}~\cite{10.1093/mnras/stu1771} obtained $Y_p = 0.2551\pm 0.0022$ from the observation of 45 extragalactic HII regions.
However, Aver, Olive and Skillman~\cite{Aver:2015iza} reanalyzed the data of Ref.~\cite{10.1093/mnras/stu1771} and obtained $Y_p= 0.2449 \pm 0.0040$.
These two measurements are inconsistent with each other.
Since more recent measurements~\cite{Fern_ndez_2018,Valerdi:2019beb} are in good agreement with result the of Ref.~\cite{Aver:2015iza}, in this paper we adopt the value obtained by  Aver, Olive and Skillman~\cite{Aver:2015iza},
\begin{equation}
     Y_p = 0.2449 \pm 0.0040.
     \label{eq:obs_const_He4_AOS}
\end{equation}
On the other hand, the Particle Data Group~\cite{Tanabashi:2018oca} adopts the more stringent constraint $Y_p=0.245\pm 0.003$.

As for the baryon-to-photon ratio $\eta_B$ we adopt
\begin{align}
  \eta_B = (6.138 \pm 0.038) \times 10^{-10},
  \label{eq:etaobs}
\end{align}
from Planck observations~\cite{Aghanim:2018eyx}.

\subsection{Constraints from BBN}
\begin{figure}[tb]
  \begin{center}
    \includegraphics[clip,width=0.49\textwidth]{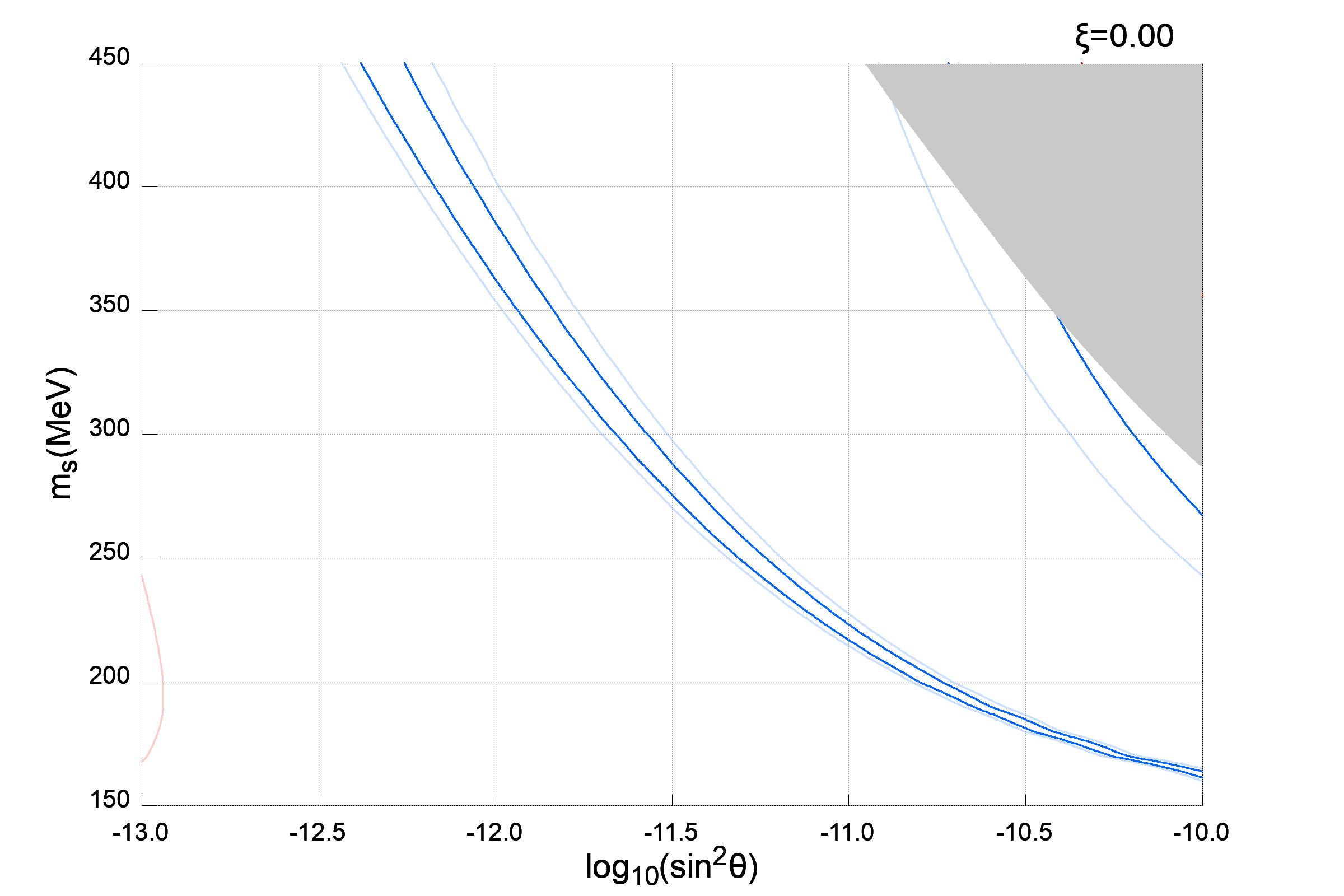}
    \includegraphics[clip,width=0.49\textwidth]{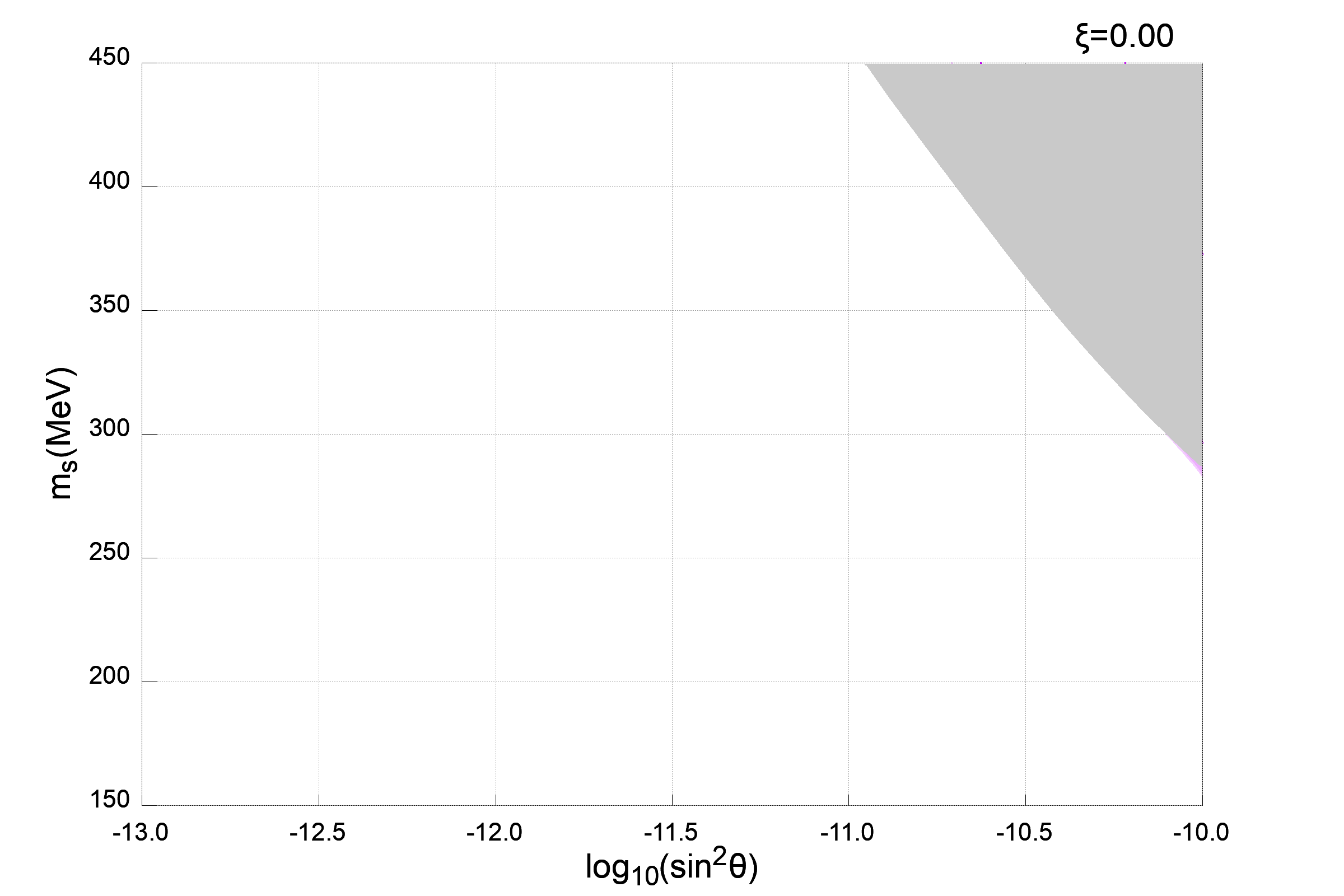}
    \caption{Regions of agreement of our numerical calculations of BBN with sterile neutrinos and data.
    In the left panel, dark and light blue lines indicate the boundaries of the regions allowed at the 68\% and 95\% C.L. by D/H data, respectively. No regions allowed by $Y_P$ limits were found.
    The right panel shows the results of the combined analysis of D/H and $Y_P$.
    There are no allowed regions in the combined analysis.
	Our calculations are not valid in the gray shaded region. All the un-shaded region is thus inconsistent with the BBN data at the 95\% C.L.}
    \label{fig:BBNresult w/o asymmetry}
  \end{center}
\end{figure}

We show the result of our BBN numerical calculations in Fig.~\ref{fig:BBNresult w/o asymmetry}.
We modified the public BBN code \texttt{PArthENoPE2.0}~\cite{Pisanti:2007hk,Consiglio:2017pot} for our purpose.
In our calculations, the baryon density after the BBN epoch is fixed by the estimation of Planck 2018~\cite{Aghanim:2018eyx}.
For the uncertainties of D/H and $Y_P$, we included the uncertainties of the \texttt{PArthENoPE2.0} calculation, the baryon density of Planck 2018, and the observational estimations.
Fig.~\ref{fig:BBNresult w/o asymmetry} shows the region where the numerical calculations and the observational estimates match at the 68\% and 95\% confidence levels. As observational estimates we used Cooke \textit{et al.}~\cite{Cooke:2017cwo} for D/H and Aver \textit{et al.} ~\cite{Aver:2015iza} for $Y_P$. In the gray shaded region our estimation for $f$ would require non-negligible corrections, as discussed in Sec.~\ref{sec:annihilation of decay product neutrino}, and we do not consider this parameter region in this paper.

As mentioned before, both the Hubble expansion speed-up effect and entropy production by the sterile neutrinos increases $^4$He.
Thus, $^4$He is overproduced unless the sterile neutrinos decay before BBN starts. On the other hand, the increase of the D abundance by the speed-up effect can be compensated by its decrease due to entropy production, which yields the narrow allowed region between blue lines for D in Fig.~\ref{fig:BBNresult w/o asymmetry}, which is consistent with the previous constraint given in Ref.~\cite{Bolton:2019pcu}.
As a result, there is no region which satisfies the observational constraints on both $^4$He and D in the unshaded region in Fig.~\ref{fig:BBNresult w/o asymmetry}.
Therefore, we conclude that the unshaded region in Fig.~\ref{fig:BBNresult w/o asymmetry} is inconsistent with the BBN data at the 95\% C.L.
This is a more severe constraint than that from $N_{\mathrm{eff}}$.

\section{BBN with lepton asymmetries}
\label{sec:sterile BBN w/ lepton asymmetry}
In the previous section we have seen that the sterile neutrinos we considered are almost ruled out by the overproduction of $^4$He.
However, the BBN constraint can be alleviated if some effect decreases the $^4$He abundance.
At the freeze-out of the weak interactions, the neutron-proton ratio in chemical equilibrium is given by
\begin{equation}
  \left.\frac{n_n}{n_p}\right|_{\mathrm{eq}} \simeq \exp\left[ -\frac{m_n-m_p+\mu_{\nu_e}}{T} \right],
\end{equation}
where $\mu_e$ is the chemical potential of the electron neutrinos with momentum distribution function $f(p)= [\exp((p-\mu_e)/T) +1]^{-1}$.
Therefore, a positive $\mu_{\nu_e}$ reduces the neutron-proton ratio at the freeze-out and can cancel the effect of the Hubble expansion speed-up by the sterile neutrinos. In the following, we assume that the three active neutrinos have the same chemical potential $\mu$ and consider the effect of a large $\xi \equiv \mu/T_{\nu}$ on BBN.

A large $\xi$ implies that there exists a large lepton asymmetry of the Universe. We assume that the lepton asymmetry is generated after sterile neutrinos decouple (if they reach thermal equilibrium) or are produced via active sterile oscillations and with a negligible accompanying entropy production, as explained in App.~\ref{sec:lepton_asymmetry}. With these assumptions the large lepton asymmetry does not affect the production of the sterile neutrinos or the subsequent evolution of its number density. Since charge neutrality requires that the number of charged leptons and anti-leptons should  be equal to the baryon asymmetry, the large lepton asymmetry we assume is carried by the three active neutrinos. The lepton asymmetry $\eta_L$, the ratio of the lepton number density to the entropy density, is
\begin{equation}
    \eta_L = \frac{n_{L}}{s} = \frac{1}{s} \sum_{i=e,\mu\tau} (n_{\nu_i}-n_{\bar{\nu}_{i}}) = 0.106~ \xi~.
\end{equation}
Here $s$ is the entropy density before the entropy production by the sterile neutrino decays.
\begin{figure}[tb]
  \begin{center}
    \includegraphics[clip,width=0.44\textwidth]{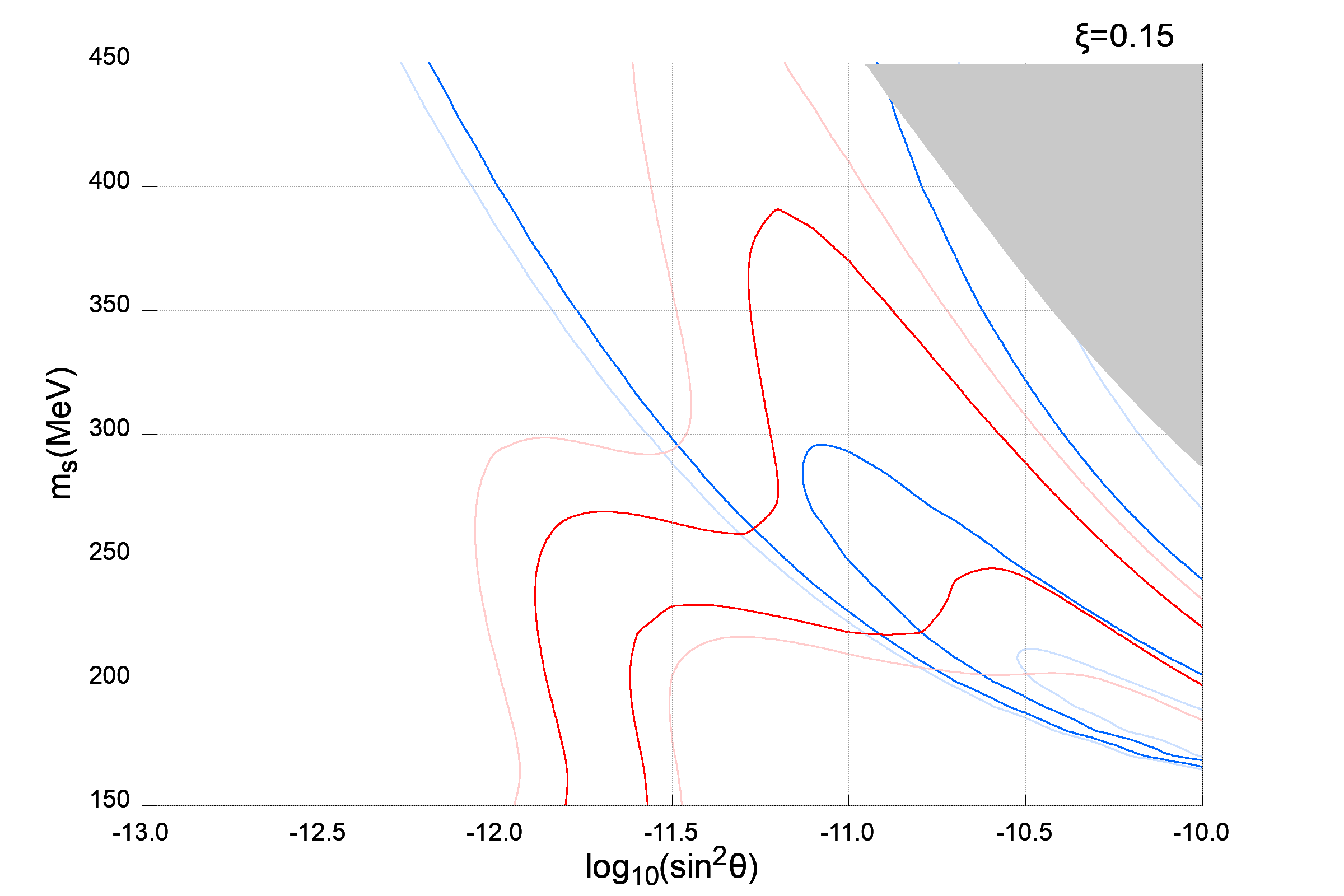}
    \includegraphics[clip,width=0.44\textwidth]{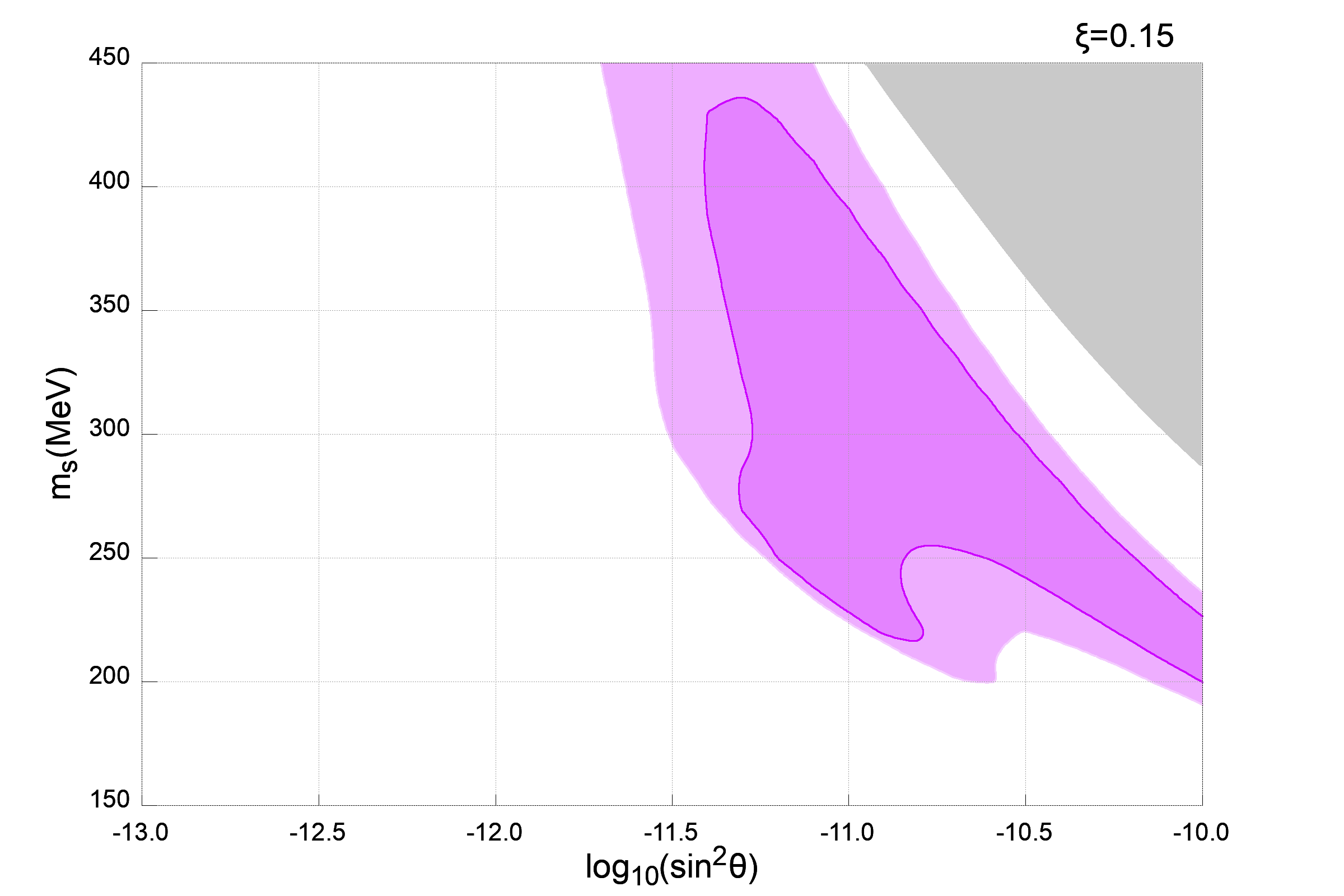}
     \caption{Results of our numerical calculation of BBN with  sterile neutrinos and a lepton asymmetry with $\xi=0.15$. In the left panel, dark and light blue lines indicate the boundaries of the regions of compatibility at the 68\% and 95\% C.L. for D/H, respectively and dark and light red lines indicate the same for $Y_P$. In the right panel, dark and light purple indicate regions allowed at the 68\% and 95\% C.L. by the combined analysis of D/H and $Y_P$.}
      \label{fig:BBNresult chi square1}
  \end{center}
\end{figure}

\begin{figure}[htbp]
  \begin{center}
    \includegraphics[clip,width=0.8\textwidth]{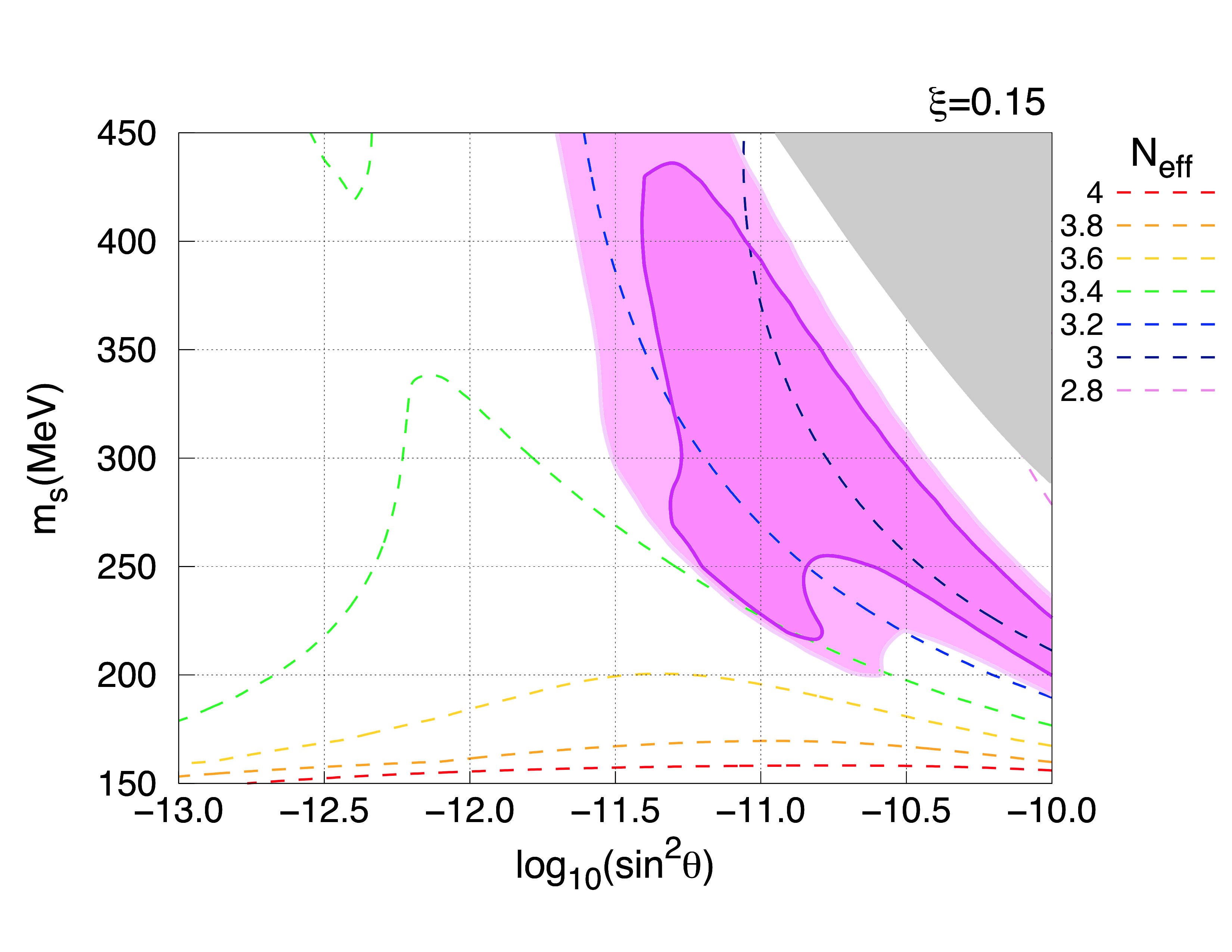}
    \caption{$N_{\mathrm{eff}}$ values and regions allowed by our BBN results with $\xi = 0.15$ (as in the righ panel of Fig. 9).
    The allowed region includes the range $N_{\mathrm{eff}} \simeq 3.2 - 3.4$, which can alleviate the tension between local and early Universe measurements of the  Hubble constant~\cite{Bernal:2016gxb}.
    }
    \label{fig:BBNresult Neff}
  \end{center}
\end{figure}

The results of the numerical calculations with $\xi = 0.15$ are shown in Fig.~\ref{fig:BBNresult chi square1}.
Results with other values of $\xi$ and comments on the choice of observational constraints are given in App.~\ref{sec:BBN result appendix}.
For $\xi \simeq 0.15$, $N_{\mathrm{eff}} \simeq 3.2 - 3.4$ can be realized while maintaining the consistency of D/H and $Y_P$ as shown in Fig.~\ref{fig:BBNresult Neff}.
This value of $N_{\mathrm{eff}}$ can alleviate the Hubble tension between CMB observations and direct measurements~\cite{Bernal:2016gxb}.

\begin{figure}[htbp]
  \begin{center}
    \includegraphics[clip,width=0.8\textwidth]{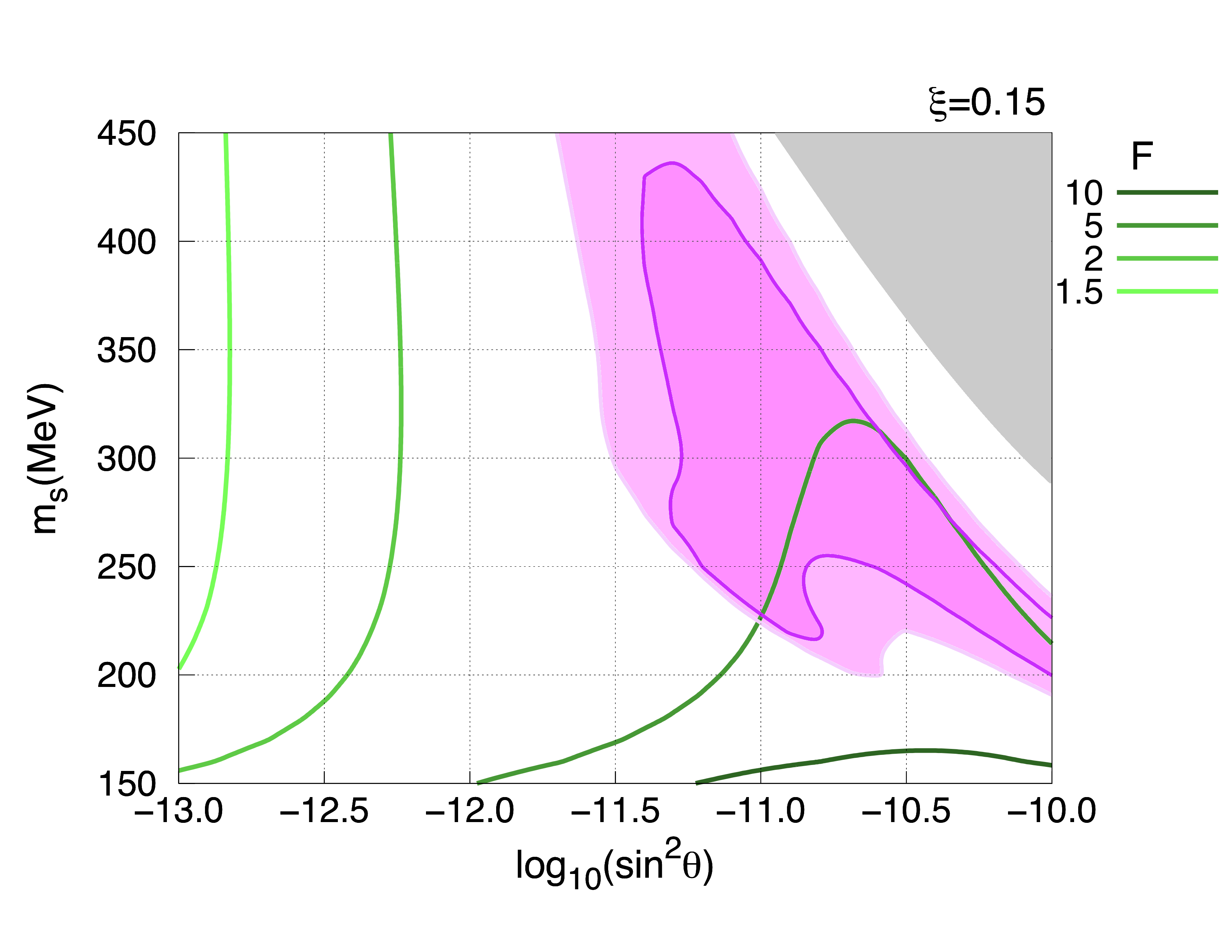}
    \caption{Ratio $F$ of the entropy after and before the  sterile neutrino decays and the region allowed by our BBN results with $\xi = 0.15$ (as in the right panel of Fig.~9). In this region
    $F \simeq 5$.
    }
    \label{fig:BBNresult 015 F}
  \end{center}
\end{figure}

 The $N_{\mathrm{eff}} \simeq 3.0 - 3.4$ range  allowed in our BBN analysis for  $\xi \simeq \mathcal{O}(0.1)$  is consistent with current constraints imposed on the $(N_{\mathrm{eff}}, \xi)$ plane by CMB and other observations~\cite{Popa:2008tb,Caramete:2013bua,Nunes:2017xon}.
Note that our scenario predicts that the value of $\xi$ relevant for CMB observations, call it $\xi_{\rm CMB}$, is actually the $\xi$ we used divided by $F$, i.e. $\xi_{\rm CMB}=\xi/F$, because the entropy generation due to the sterile neutrinos decay also dilutes the lepton asymmetry.
In Fig.~\ref{fig:BBNresult 015 F}, we show lines of constant entropy ratio $F$ and the BBN allowed region for $\xi = 0.15$. We can see that the region consistent with BBN indicates that $F \simeq 5$.

\begin{figure}[htbp]
  \begin{center}
    \includegraphics[clip,width=0.49\textwidth]{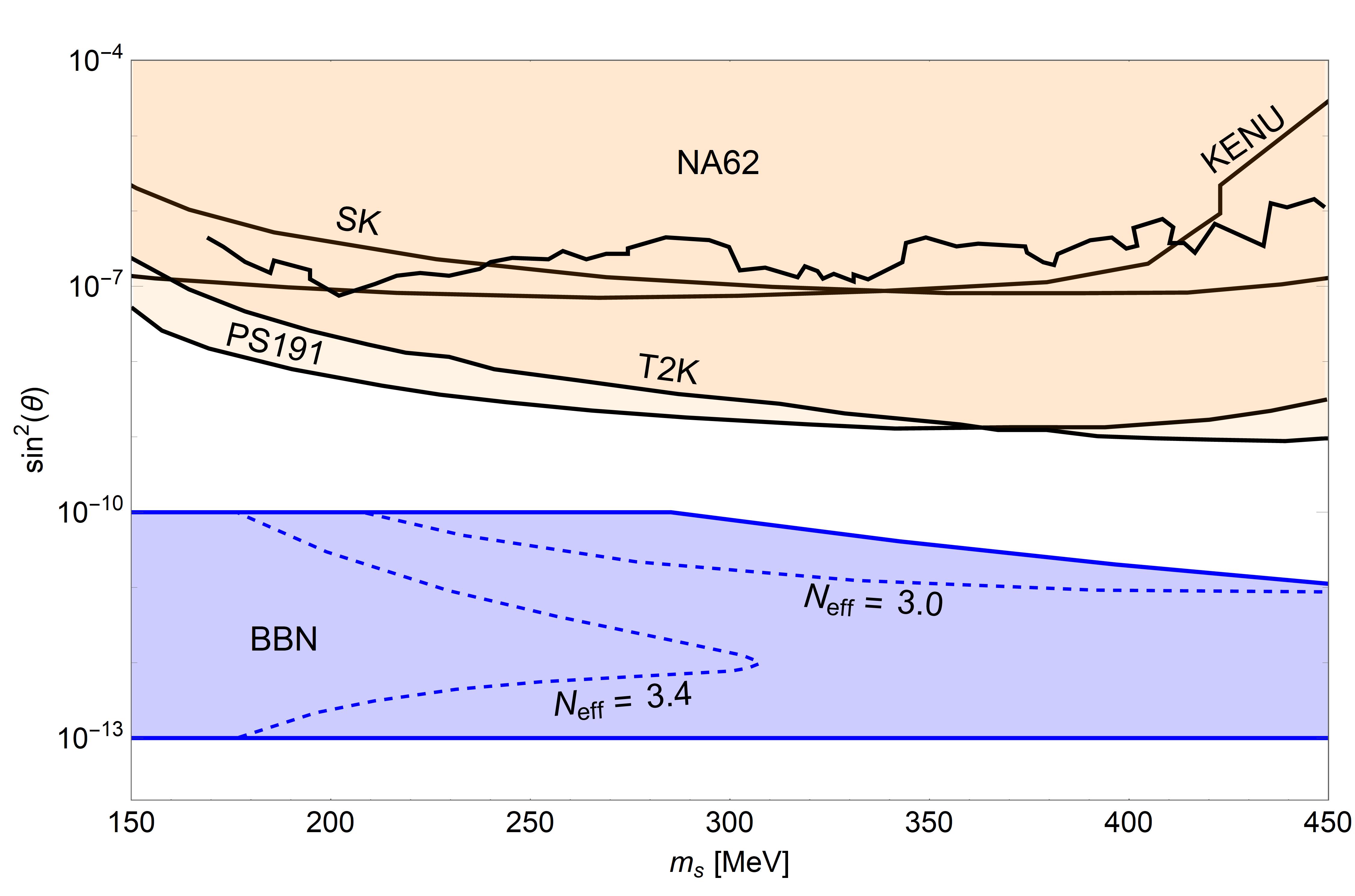}
    \includegraphics[clip,width=0.49\textwidth]{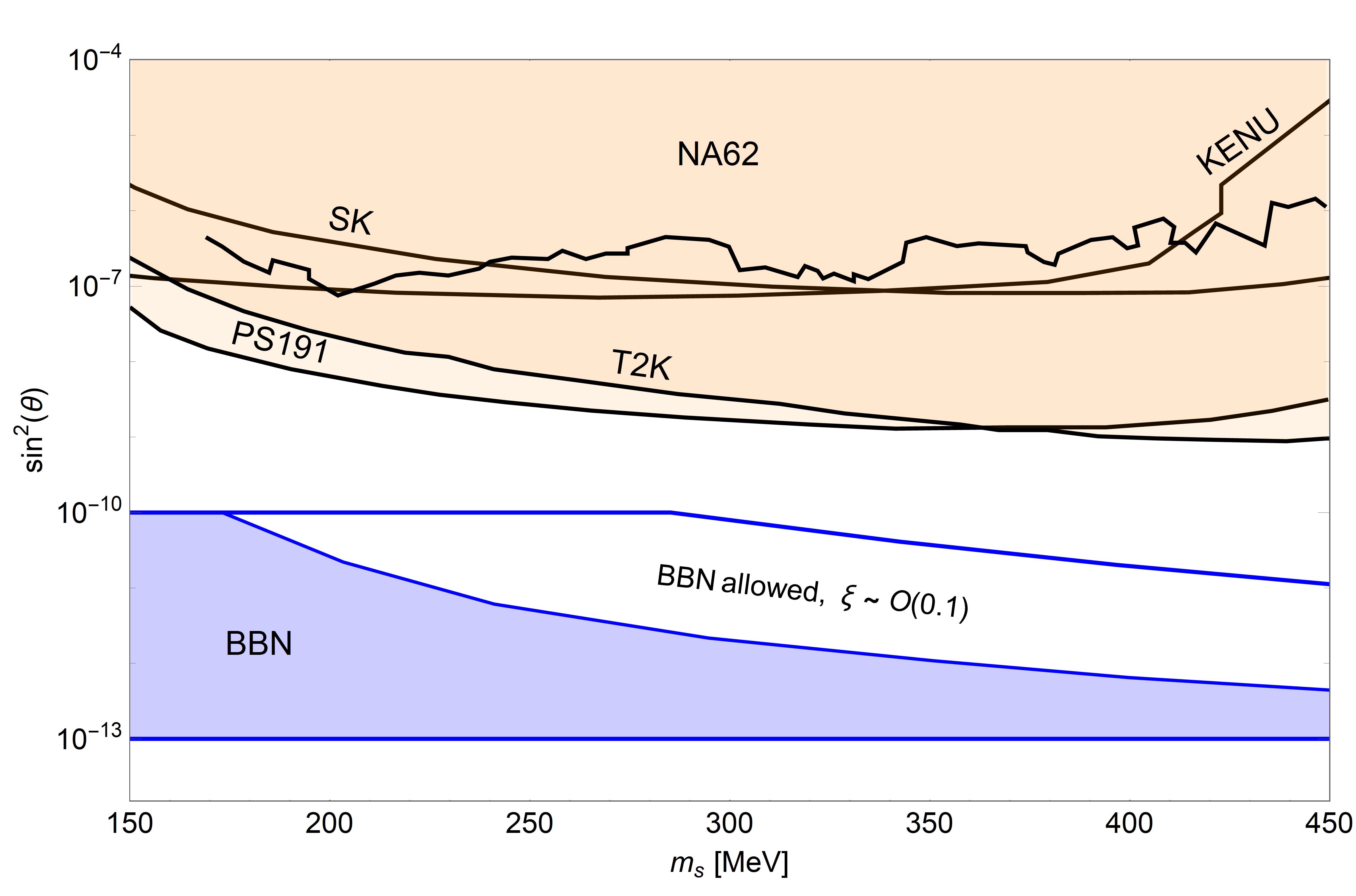}
    \caption{Regions inconsistent with the BBN data (blue) of heavy sterile neutrinos mixing with electron neutrinos for zero lepton asymmetry [left panel] and large lepton asymmetry with $\xi = \mathcal{O}(0.1)$ [right panel], along with regions (orange) rejected by existing upper bounds from the T2K near detector~\cite{Abe:2019kgx}, NA62~\cite{CortinaGil:2017mqf}, PS191~\cite{Bernardi:1985ny,Bernardi:1987ek}, Super-Kamiokande (labeled ``SK'') - as derived in Ref.~\cite{Coloma:2019htx,Kusenko:2004qc}, and precision measurements of kaon decays (labeled ``KENU'')~\cite{Lazzeroni:2012cx} as derived in Ref.~\cite{Bryman:2019bjg}. The (dark blue) upper boundary of the region in which our BBN results apply corresponds to $\sin^2 (\theta)= 10^{-10}$ and to the lower boundary of the gray region shown in previous figures, excluded from our analysis (because the energy fraction $f$ injected into the plasma would receive non-negligible corrections as, discussed in text). The (dark blue) lower boundary is set by the reach of our numerical calculations, $\sin^2 (\theta)= 10^{-13}$. The lines of $N_{\rm eff} = 3.0$ and $N_{\rm eff} = 3.4$ for no lepton asymmetry are indicated (dashed blue lines) in the left panel. The right panel shows that a sizable lepton asymmetry, $\xi = \mathcal{O}(0.1)$, allows to remove BBN constraints in the part of the region where $N_{\rm eff} = 3.0 -3.4$, which is consistent with CMB observations and alleviates the Hubble constant tension.
    \label{fig:full BBN constraint}}
  \end{center}
\end{figure}

In Fig.~\ref{fig:full BBN constraint} we display the regions of mass and mixing inconsistent with the BBN data, with and without lepton asymmetry, as well as the current laboratory constraints on a heavy sterile neutrino that mixes with the electron neutrino.

\section{Discussion}
\label{sec:result}
In this work, we investigated the impact on BBN of heavy  sterile neutrinos with the mass range $150~\mathrm{MeV} < m_s < 450~\mathrm{MeV}$ that decay after electron-positron annihilation. We find that BBN severely constrains the sterile neutrino mass $m_s$ and active-sterile mixing $\sin \theta$ parameter space. The obtained constraints are significantly more stringent than those obtained from limits on $N_{\mathrm{eff}}$.

We found that the constraints from BBN can be drastically relaxed by the introduction of a large lepton asymmetry $\xi \sim \mathcal{O}(0.1)$\footnote{During the preparation of this manuscript, a paper on a related topic was released~\cite{Grohs:2020xxd}. This paper suggests that the combination of entropy generation, extra radiation energy, and lepton asymmetries can be consistent with BBN.  Our work confirms this suggestion by calculating BBN in a specific situation.}, assuming the large lepton asymmetry is accounted for independently of the sterile neutrino production. Notice that such a large lepton asymmetry should be produced at sufficiently late times. If a large lepton asymmetry is produced in the early Universe at $T > 100$~GeV, sphaleron processes convert a significant fraction of the lepton number to baryon number, which leads to a too large baryon asymmetry $\eta_B \gg 10^{-9}$ of the present Universe.
In order to produce a large lepton asymmetry with $\xi\sim\mathcal{O}(0.1)$, a special setup is required,
for example if Q-balls that contain lepton charges form in the early Universe and decay after the electroweak phase transition.

For simplicity, we assumed that the large lepton asymmetry is generated at $T<$~few~GeV,  after the decoupling (if they reached thermal equilibrium) or production (via active-sterile oscillations) of sterile neutrinos, thus these processes are not affected by it,
as mentioned in App.~\ref{sec:sterile BBN w/ lepton asymmetry}. Moreover, as explained in App.~\ref{sec:lepton_asymmetry}, the entropy production associated with the generation of the large lepton number asymmetry via Q-balls decay  (which is model-dependent) could be negligible, as we assume here. With these assumptions the production and evolution of the sterile neutrino number density is the same as without the large lepton density.

With a large lepton asymmetry $\xi \sim \mathcal{O}(0.1)$, BBN limits allow the $N_{\mathrm{eff}} \simeq 3.2 - 3.4$ range, that could alleviate the tension between local and early Universe measurements of the Hubble constant. The values of $N_{\mathrm{eff}}$ and $\xi$ allowed by our BBN analysis are consistent with CMB limits.
 The relaxation of the BBN limits on $N_{\mathrm{eff}}$ with $\xi \sim \mathcal{O}(0.1)$ is shown in Fig.~\ref{fig:full BBN constraint}, where we display the regions of mass and mixing inconsistent with the BBN data with and without a lepton asymmetry, as well as current laboratory upper bounds on the mixing, assuming that the sterile neutrino only mixes with the electron neutrino (as we do in this work).

The cosmological effects of heavy decaying sterile neutrinos go beyond just BBN. Non-thermal active neutrinos with high energy produced in the decay of the sterile neutrinos have a larger free-streaming length than the thermal neutrinos in the standard scenario. Therefore, the decay of the sterile neutrinos induces a suppression of the matter power spectrum at large scales (see e.g. Ref.~\cite{Gelmini:2019deq}).
Furthermore, while we have concentrated on phenomenological aspects related to BBN, as already noted in Ref.~\cite{Fuller:2011qy}, heavy decaying sterile neutrinos provide an interesting setting for model-building. In particular, the significant entropy dilution  due to the sterile neutrino decays can allow light decoupled particles to avoid cosmological bounds, and the change in the baryon-to-entropy ratio it produces affects models of baryogenesis.

The parameter region studied in this work is restricted by some assumptions and the reach of our numerical calculations. By expanding the explored regions, we may be able to impose more severe constraints on sterile neutrinos or find more extended allowed regions, which will be done in future work.

\section*{Acknowledgments}
M. Kawasaki is grateful to the Department of Physics and Astronomy, University of California, Los Angeles for hospitality during the time when this work started.
This work was supported by JSPS KAKENHI Grant Nos. 17H01131 (M.K.) and 17K05434 (M.K.), MEXT KAKENHI Grant Nos. 15H05889 (M.K.), World Premier International Research Center Initiative (WPI Initiative), MEXT, Japan (M. K. and K. M.), the Program of Excellence in Photon Science (K. M.), and the JSPS Research Fellowships for Young Scientists Grant No. 20J20248 (K. M.). The work of G.G., A.K and V.T. was supported by the U.S. Department of Energy (DOE) Grant No. DE-SC0009937.

\appendix
\section{Decay channels of sterile neutrino}
\label{sec:decay channel}
In the following we summarize the decay channels of a sterile neutrino with the mass of $150~\mathrm{MeV} \lesssim m_s \lesssim 450~\mathrm{MeV}$.
Here we assume that the sterile neutrino mixes only with the electron neutrinos and show only the channels where the sterile neutrino changes into an electron neutrino.
If the sterile neutrino is Majorana, each channel has its counterpart where the sterile neutrino changes into an anti-electron neutrino. In this paper, we assume that the sterile neutrino is Majorana and the decay rate is double the sum of the decay rates below. For each decay channel, we show the decay rate and the fraction of the sterile neutrino rest mass deposited in the thermal plasma $f$.
In estimating $f$ we assume that neutrinos in the final state do not interact with the other particles and do not give their energy to the thermal plasma. The
justification of this assumption is examined in Appendix~\ref{sec:annihilation detail}. In the calculation, the physical constants were taken from~\cite{Tanabashi:2018oca}. The decay rates in Sec.~\ref{sec:3nu channel} to Sec.~\ref{sec:epi channel} follow~Ref.~\cite{Gorbunov:2007ak} and that in Sec.~\ref{sec:gamma channel} follows~Ref.~\cite{Abazajian:2001nj}. The discussion on $f$ follows~Ref.~\cite{Fuller:2011qy}.

\subsection{\texorpdfstring{$\nu_s \to 3\nu$}{}}
\label{sec:3nu channel}
The decay rate for this channel is
\begin{equation}\label{eq:3nu rate}
  \Gamma_{3\nu} = \frac{G_F^2}{192\pi^3}m_s^5\sin^2\theta,
\end{equation}
where $G_F$ is the Fermi constant, $m_s$ is the mass of the sterile neutrino, and $\theta$ is the mixing angle between $\nu_e$ and $\nu_s$.
The neutrinos in the final state are decoupled from the thermal bath, then $f=0$.

\subsection{\texorpdfstring{$\nu_s \to \nu_e + e^+ + e^-$}{}}
\label{sec:nuee channel}
The decay rate for this channel is
\begin{align}
  \Gamma_{\nu e^- e^+}
  &= \frac{G_F^2}{192\pi^3} m_s^5 \sin^2\theta \nonumber\\
  \times&\left[
    \frac{1+4\sin^2\theta_w + 8\sin^4\theta_w}{4}
    \left(
      (1-14x_e^2 - 2x_e^4 - 12x_e^6)\sqrt{1-4x_e^4}
      + 12x_e^4(x_e^4 -1)L_e)
    \right)\right.\nonumber\\
    &\left.+
    2\sin^2\theta_w(2\sin^2\theta_w+1)
    \left(
      x_e^2(2+10x_e^2-12x_e^4)\sqrt{1-4x_e^4}
      + 6x_e^4(1-2x_e^2+2x_e^4)L_e
    \right)
  \right],
\end{align}
where
\begin{equation}
  x_i \equiv \frac{m_i}{m_s},\quad
  L_i \equiv \log
  \left[
    \frac{1-3x_i^2-(1-x_i^2)\sqrt{1-4x_i^2}}{x_i^2(1+\sqrt{1-4x_i^4})}
  \right],\quad
  (i = e , \mu)
\end{equation}
and $\theta_w$ is the Weinberg angle.
In this channel, the sterile neutrino mass is equally distributed to the three final state particles, which leads to $f \sim 2/3$
\subsection{\texorpdfstring{$\nu_s \to \nu_e + \mu^- + \mu^+$}{}}
\label{sec:numumu channel}
The decay rate for this channel is
\begin{align}
  \Gamma_{\nu \mu^- \mu^+}
  &= \frac{G_F^2}{192\pi^3} m_s^5 \sin^2\theta \nonumber\\
  \times&\left[
    \frac{1 - 4\sin^2\theta_w + 8\sin^4\theta_w}{4}
    \left(
      (1 - 14x_\mu^2 - 2x_\mu^4 - 12x_\mu^6)\sqrt{1 - 4x_\mu^4}
      + 12x_\mu^4(x_\mu^4 - 1)L_\mu)
    \right)\right.\nonumber\\
    &\left.+
    2\sin^2\theta_w(2\sin^2\theta_w - 1)
    \left(
      x_\mu^2(2 + 10x_\mu^2-12x_\mu^4)\sqrt{1 - 4x_\mu^4}
      + 6x_\mu^4(1 - 2x_\mu^2 + 2x_\mu^4)L_\mu
    \right)
  \right].
\end{align}
This channel is available if $m_s \gtrsim 2m_\mu$.
The produced muons decay mainly with $\mu^- \to e^- + \nu_\mu + \bar{\nu}_e$.
In this process, each (anti-)neutrino has average energy $E_{\nu\mu} = 34.33~\mathrm{MeV}$.
Therefore, $f \sim 2/3 - 4 E_{\nu\mu}/m_s$ when $m_s \gg 2m_{\mu}$.
\subsection{\texorpdfstring{$\nu_s \to e^- + \nu_{\mu} + \mu^+$}{}}
\label{sec:enumu channel}
The decay rate for this channel is
\begin{equation}\label{eq:enumu rate}
  \Gamma_{\nu_e\mu^-\mu^+} = \frac{G_F^2}{192\pi^3} m_s^5 \sin^2\theta
  (1 - 8x_{\mu}^2 + 8x_{\mu}^6 - x_{\mu}^8 - 12x_{\mu}^4\log x_{\mu}^2)
\end{equation}
This channel is available if $m_s \gtrsim m_e + m_{\mu}$.
Like Sec.~\ref{sec:numumu channel}, the muon decay take some energy away from the plasma and $f \sim 2/3 - 2 E_{\nu\mu}/m_s$ when $m_s \gg m_{e} + m_{\mu}$.
\subsection{\texorpdfstring{$\nu_s \to \nu_e + \pi^0$}{}}
\label{sec:nupi channel}
The decay rate for this channel is
\begin{equation}\label{eq:nupi rate}
  \Gamma_{\nu_e \pi^0} = \frac{G_F^2f_{\pi}^2}{32\pi}m_s^3
  \left(
    1-\frac{m^2_{\pi^0}}{m^2_s}
  \right)^2,
\end{equation}
where $m_{\pi^0}$ is the neutral pion mass and $f_{\pi} = 130.2~\mathrm{MeV}$ is the pion decay constant.
This channel is available if $m_s \gtrsim m_{\pi^0}$.
The kinetic energy that the produced neutrino takes away is $(m_s^2-m_{\pi^0}^2)/2m_s$ and $f \sim 1/2+m_{\pi^0}^2/2m_s^2$.
\subsection{\texorpdfstring{$\nu_s \to e^- + \pi^+$}{}}
\label{sec:epi channel}
The decay rate for this channel is
\begin{align}\label{eq:epi rate}
  \Gamma_{e^- \pi^+} &= \frac{G_F^2f_{\pi}^2}{16\pi}
  \left|V_{ud}\right|m_s^3 \sin^2 \theta
  \left(
    \left(
      1-\frac{m^2_e}{m^2_s}
    \right)^2
    -
    \frac{m_{\pi^+}^2}{m_s^2}
    \left(
      1+\frac{m_e^2}{m_s^2}
    \right)
  \right)\nonumber\\
  &\times \sqrt{
    \left(
      1-\frac{(m_{\pi^+}-m_e)^2}{m_s^2}
    \right)
    \left(
      1-\frac{(m_{\pi^+}+m_e)^2}{m_s^2}
    \right)
  },
\end{align}
where $V_{ud}$ is the CKM matrix element.
This channel is available if $m_s \gtrsim m_{\pi^0}$.
The produced charged pion decays mainly with $\pi^+ \to \mu^+ + \nu_{\mu}$ and the $\nu_{\mu}$ has average energy $E_{\nu\pi} = 29.79~\mathrm{MeV}$.
The timescale of this decay is much shorter than the scattering and dynamical timescale,
thus we can consider that this decay occurs instantaneously.
The produced muon then decays into $e^-$, $\nu_{\mu}$, and $\bar{\nu}_e$.
As a result, the average energy taken away to the decoupled neutrino sea is $\langle E_{\mathrm{dec}} \rangle = E_{\nu\pi} + 2E_{\nu\mu} + (5/6) E_{\pi\mathrm{KE}}$,
where the kinetic energy of the pion is $E_{\pi\mathrm{KE}} = m_s - E_e - m_{\pi^+}$ and the electron energy is $E_e = (m_s^2 - m_{\pi^+}^2 + m_e^2) / 2m_s$.
Therefore, $f \sim 1 - (E_{\nu\pi} + 2E_{\nu\mu} + (5/6) E_{\pi\mathrm{KE}}) / m_s$.
\subsection{\texorpdfstring{$\nu_s \to \nu_e + \gamma$}{}}
\label{sec:gamma channel}
The decay rate for this channel is
\begin{equation}\label{eq:gamma rate}
  \Gamma_{\nu\gamma} = \frac{9G_F^2}{512\pi^4}\alpha m_s^5 \sin^2\theta,
\end{equation}
where $\alpha$ is the fine structure constant.
Because $\nu_s$ is much heavier than $\nu_e$ and $\gamma$, the rest mass of the sterile neutrino is equally distributed into the two particles.
Therefore, $f \sim 1/2$.

\section{Annihilation of high-energy neutrinos from the decay}
\label{sec:annihilation detail}
Here we consider the correction to $\overline{f}$ due to the scatterings and annihilations of the high energy active neutrinos produced from the sterile neutrino decays.
While scatterings and annihilations have cross sections of similar magnitude,
annihilation processes such as $\nu_e + \overline{\nu}_e \to e^- + e^+$ inject the maximal energy into the plasma.
Therefore we focus on annihilation processes in order to evaluate the correction to $\overline{f}$.
In addition, the decay process most relevant to the correction to $\overline{f}$ is expected to emit high energy neutrino(s) and have a measurable branching ratio.
From this point of view, we focus on the $\nu_s \to \nu_e + \pi^0$ channel.
\subsection{Lorentz transformation of cross sections}
First, we discuss the cross section.
Let us consider the two massless particle that form the angle $\alpha$.
The momenta of these particle can be written as
\begin{equation}
  p_A^{\mu} = E_A\left(
  \begin{array}{c}
    1 \\
    \sin \alpha/2 \\
    \cos \alpha/2 \\
    0
  \end{array}
  \right),\quad
  p_B^{\mu} = E_B\left(
  \begin{array}{c}
    1 \\
    -\sin \alpha/2 \\
    \cos \alpha/2 \\
    0
  \end{array}
  \right).
\end{equation}
By the Lorentz boost along the $y$ axis, these momenta can be transformed to
\begin{equation}
  p_{A'}^{\mu} = E_A \sin \alpha/2\left(
  \begin{array}{c}
    1 \\
    1 \\
    0 \\
    0
  \end{array}
  \right),\quad
  p_{B'}^{\mu} = E_B \sin \alpha/2\left(
  \begin{array}{c}
    1 \\
    -1 \\
    0 \\
    0
  \end{array}
  \right)
\end{equation}
and by making the Lorentz boost along the $x$ axis successively, we obtain the momenta in the center of mass frame,
\begin{equation}
  p_{A''}^{\mu} = \sqrt{E_A E_B} \sin \alpha/2\left(
  \begin{array}{c}
    1 \\
    1 \\
    0 \\
    0
  \end{array}
  \right),\quad
  p_{B''}^{\mu} = \sqrt{E_A E_B} \sin \alpha/2\left(
  \begin{array}{c}
    1 \\
    -1 \\
    0 \\
    0
  \end{array}
  \right).
\end{equation}
Since the cross sections of two-body scatterings are given by
\begin{equation}
  \mathrm{d}\sigma =
  \frac{1}{2E_A 2E_B|v_A - v_B|}\left(
  \prod_f \frac{\mathrm{d}^3p_f}{(2\pi)^3}\right)
  \frac{1}{2E_f}|\mathcal{M}|^2(2\pi)^4\delta^{(4)}(p_A+p_B-\sum p_f),
\end{equation}
the dependence of the cross sections on frames is represented by $L \equiv 1/(E_A E_B|v_A - v_B|)$.
$L=(2 E_A E_B \sin \alpha/2)^{-1}$ in the first frame and $L''=(E_A E_B \sin^2 \alpha/2)^{-1}$ in the center of mass frame.
Therefore, we can obtain the relation between the cross section in the first frame $\sigma$ and that in the center of mass frame $\sigma_{\mathrm{CM}}$ as
\begin{equation}
  \sigma = \frac{L}{L''}\sigma_{\mathrm{CM}} = \sigma_{\mathrm{CM}} \sin \frac{\alpha}{2}.
  \label{eq:cross section LT}
\end{equation}
\subsection{Annihilation of the decay products with the background neutrinos}
Next we consider the annihilation of the decay products with the background neutrinos.
In the following, we calculate the averaged probability $P_{\mathrm{ann,BG}}$ that the neutrinos from the sterile neutrino decays annihilate with the background neutrinos.
In the following we consider the cross sections in the center of mass frame of each process and then move to the rest frame of the radiation bath.

The amplitude squared $|\mathcal{M}|^2$ of $\nu_e(p_1) + \overline{\nu}_e(p_2) \to e^-(p_3) + e^+(p_4)$ is
\begin{equation}
  |\mathcal{M}|^2 = 32G_F^2\left[
    (C_V-C_A)^2(p_1\cdot p_3)^2 + (C_V + C_A)^2 (p_1\cdot p_4)^2 +(C_V^2-C_A^2)m_e^2 p_1\cdot p_2
  \right],
\end{equation}
where $C_V = -0.5+2\sin^2 \theta_W$, $C_A = -0.5$, and $\theta_W$ is the Weinberg angle.
In our cases, since the momenta in the center of mass frame are much larger than $m_e$, we ignore the term with $m_e^2$.
By defining the angle made by $\vec{p}_1$ and $\vec{p}_3$ as $\theta$, $|\mathcal{M}|^2$ can be rewritten as
\begin{equation}
  |\mathcal{M}|^2 = 32G_F^2E_1^4\left[
    (C_V-C_A)^2(1-\cos \theta)^2 + (C_V + C_A)^2 (1+\cos\theta)^2
  \right],
\end{equation}
where $E_1$ is the zero component of $p_1^{\mu}$.
In the center of mass frame, the cross section of $2 \to 2$ process is given by
\begin{equation}
  \left(\frac{\mathrm{d}\sigma}{\mathrm{d}\Omega}\right)_{\mathrm{CM}} =
  \frac{|\mathcal{M}|^2}{64\pi^2 (2E_1)^2} =
  \frac{G_F^2E_1^2}{8\pi^2}\left[
    (C_V-C_A)^2(1-\cos \theta)^2 + (C_V + C_A)^2 (1+\cos\theta)^2
  \right],
\end{equation}
where $\Omega$ is the solid angle of the final electron.
By integrating over $\Omega$, the total cross section is obtained as
\begin{equation}
  \sigma_{\mathrm{CM}} = \frac{4G_F^2 E_1^2}{3\pi}(C_V^2 + C_A^2).
\end{equation}

In the rest frame of the bath, we define the energy of the active neutrino produced in the sterile neutrino decay and the background neutrino as $E_A$ and $E_B$.
Then $E_1$ can be written in terms of $E_A$ and $E_B$ as
\begin{equation}
  E_1 = \sqrt{E_A E_B}\sin \frac{\alpha}{2},
\end{equation}
and the cross section in the rest frame is
\begin{equation}
  \sigma = \frac{4G_F^2 E_A E_B}{3\pi}(C_V^2 + C_A^2)\sin^3 \frac{\alpha}{2}
\end{equation}
from (\ref{eq:cross section LT}).
The cross section averaged over $\alpha$ can be obtained by integrating over $\cos \alpha$ as
\begin{equation}
    \langle \sigma \rangle_{\alpha} =
    \frac{4G_F^2 E_A E_B}{3\pi}(C_V^2 + C_A^2) \int_{-1} ^1 \frac{\mathrm{d}\cos \alpha}{2} \sin^3 \frac{\alpha}{2}
    = \frac{8G_F^2 E_A E_B}{15\pi}(C_V^2 + C_A^2).
\end{equation}

Let us consider a sterile neutrino that decays at the time $t_d$ and emits a neutrino with an energy $E_{\nu}$ and
calculate the probability $\tilde{P}_{\mathrm{ann,BG}}(t_d)$ that this neutrino annihilates with a background neutrino.
Since the annihilation rate $\Gamma_{\mathrm{ann,BG}} = \langle \sigma \rangle n_{\nu,\mathrm{BG}}$ is proportional to the energies of neutrinos and the number density of background neutrinos,
we can normalize the annihilation rate by using that just after the sterile neutrino decay at $t_1$, $\Gamma_{\mathrm{ann,BG}}(t_1)$ is
\begin{equation}
  \Gamma_{\mathrm{ann,BG}}(t) =
  \left( \frac{a(t)}{a(t_d)} \right)^{-1}
  \left( \frac{a(t)}{a(t_1)} \right)^{-4}
  \Gamma_{\mathrm{ann,BG}}(t_1),
\end{equation}
where the first ratio of scale factors indicates the redshift of the neutrino produced in the decay and the second ratio of scale factors indicates the redshift of the background neutrinos.
Here we take $t_1$ as the time when $\Gamma_{\mathrm{decay}} = H(t_1)$.
Therefore,
\begin{align}
  \tilde{P}_{\mathrm{ann,BG}}(t_d)
  &=
  1-\exp \left[ -\int^{\infty}_{t_d} \mathrm{d}t \Gamma_{\mathrm{ann,BG}}(t) \right]\nonumber\\
  &=
  1-\exp \left[ -\int^{\infty}_{t_d} \mathrm{d}t
  \left( \frac{a(t)}{a(t_d)} \right)^{-1}
  \left( \frac{a(t)}{a(t_1)} \right)^{-4}
  \Gamma_{\mathrm{ann,BG}}(t_1) \right].
\end{align}
In order to obtain the averaged probability $P_{\mathrm{ann,BG}}$ that the neutrinos produced in the sterile neutrino decays annihilate with the background neutrinos, we convolute $ \tilde{P}_{\mathrm{ann,BG}}(t_d)$ with the time distribution of the sterile neutrino decays $\Gamma_{\mathrm{decay}}\exp(-\Gamma_{\mathrm{decay}}t_d)$:
\begin{align}
  P_{\mathrm{ann,BG}} &= \int^{\infty}_{0} \mathrm{d}t_d\Gamma_{\mathrm{decay}}e^{-\Gamma_{\mathrm{decay}}t_d}\tilde{P}_{\mathrm{ann,BG}}(t_d) \nonumber \\
  &=
  \int^{\infty}_{0} \mathrm{d}t_d\Gamma_{\mathrm{decay}}e^{-\Gamma_{\mathrm{decay}}t_d}
  \left( 1-\exp \left[ -\int^{\infty}_{t_d} \mathrm{d}t
  \left( \frac{a(t)}{a(t_d)} \right)^{-1}
  \left( \frac{a(t)}{a(t_1)} \right)^{-4}
  \Gamma_{\mathrm{ann,BG}}(t_1) \right] \right).
\end{align}
For numerical evaluations, we assume the scale factor is proportional to a power of $t$: $a\propto t^s$.
Then $H(t) = s/t$ and $\Gamma_{\mathrm{decay}} = s/t_1$ follow.
By using these equations, we can rewrite $P_{\mathrm{ann,BG}}$ as
\begin{align}
  P_{\mathrm{ann,BG}} &=
  s\int^{\infty}_{0} \mathrm{d}x_d\, e^{-s x_d}
  \left( 1-\exp \left[ -sx_d^s\frac{\Gamma_{\mathrm{ann,BG}}(t_1)}{\Gamma_{\mathrm{decay}}}
  \int^{\infty}_{x_d} \mathrm{d}x\, x^{-5s} \right] \right)\nonumber\\
  &=
  s\int^{\infty}_{0} \mathrm{d}x_d\, e^{-s x_d}
  \left( 1-\exp \left[ -s\frac{x_d^{-4s+1}}{5s-1}
  \frac{\Gamma_{\mathrm{ann,BG}}(t_1)}{\Gamma_{\mathrm{decay}}}
   \right] \right)
   \label{eq:prob ann BG}
\end{align}

Finally, let us calculate $\Gamma_{\mathrm{ann,BG}}(t_1)$.
At the time $t_1$, the number density of the background neutrinos is
\begin{equation}
  n_{\nu}(t_1) = \frac{3\zeta(3)}{4\pi^2}T_{\nu}^3(t_1)
\end{equation}
and the cross section is
\begin{equation}
  \langle \sigma \rangle = \frac{8G_F^2 E_{\nu}\cdot 3.15T_{\nu}(t_1)}{15\pi}(C_V^2 + C_A^2),
\end{equation}
where $3.15T_{\nu}(t_1)$ is the thermal averaged energy of the background neutrino.
Therefore
\begin{equation}
  \Gamma_{\mathrm{ann,BG}}(t_1) =
  \frac{2\cdot3.15\zeta(3)G_F^2 E_{\nu}}{5\pi^3}\left( \frac{4}{11} \right)^{4/3}(C_V^2 + C_A^2)T_{\gamma}^4(t_1)
  \label{eq:ann rate BG}
\end{equation}
where we used $T_{\nu} = (4/11)^{1/3}T_{\gamma}$.
$T_{\gamma}(t_1)$ can be obtained from the Friedmann equation and the condition $H(t_1) = \Gamma_{\mathrm{decay}}$:
\begin{equation}
  H(t_1) = \sqrt{\frac{\rho_{\nu_s}+\rho_R}{3M_{\mathrm{Pl}}^2}}
  = \frac{1}{\sqrt{3}M_{\mathrm{Pl}}}
  \left[
    m_s n_{\nu_s} \left( T_{\gamma}(t_1) \right)
   +
    g_* \frac{\pi^2}{30}T_{\gamma}^4(t_1)
  \right]^{\frac{1}{2}} = \Gamma_{\mathrm{decay}}
  \label{eq:decay temperature of sterile neutrino}
\end{equation}

By using (\ref{eq:prob ann BG}), (\ref{eq:ann rate BG}), and (\ref{eq:decay temperature of sterile neutrino}) and focusing on $\nu_e + \pi^0$ channel,
we can calculate $P_{\mathrm{ann,BG}}$ and the contribution of this process to the correction to $\overline{f}$, $\Delta \overline{f}_{\mathrm{ann,BG},\nu_e \pi^0}$ from
\begin{equation}
  \Delta \overline{f}_{\mathrm{ann,BG},\nu_e\pi^0} = \left(\frac{1}{2} - \frac{m_{\pi^0}^2}{2m_s^2}\right)
  P_{\mathrm{ann,BG}} B_{\nu_e\pi^0},
\end{equation}
where $B_{\nu_e\pi^0}\equiv\Gamma_{\nu_e\pi^0}/\Gamma_{\mathrm{decay}}$ is the branching ratio of
$\nu_s \to \nu_e + \pi^0$.

\subsection{Annihilation of the decay products with themselves}
Annihilations of the decay products can be treated in a similar way.
First we derive the probability $\tilde{P}_{\mathrm{ann,D}}(t_d)$ that the neutrino produced at $t_d$ annihilates. In this case, we have to integrate over the time $\tilde{t}$ when the other annihilating  neutrino is produced.
Then we can represent the annihilation rate at $t$ as
\begin{equation}
  \Gamma_{\mathrm{ann,D}}(t) =
  \tilde{\Gamma}_{\mathrm{ann,D}}(t_1)
  \left( \frac{a(t)}{a(t_d)} \right)^{-1}
  \left( \frac{a(t)}{a(t_1)} \right)^{-3}
  \int_0^t\mathrm{d}\tilde{t}\, \Gamma_{\mathrm{decay}}e^{-\Gamma_{\mathrm{decay}}\tilde{t}}
  \left( \frac{a(t)}{a(\tilde{t})} \right)^{-1},
\end{equation}
where $\tilde{\Gamma}_{\mathrm{ann,D}}(t_1)$ is the annihilation rate at $t_1$ for the neutrino produced at $t_1$ when we assume that all the sterile neutrinos instantaneously decay at $t_1$.
From a similar discussion, we obtain the averaged probability that the active neutrinos produced in the sterile neutrino decays annihilate with each other $P_{\mathrm{ann,D}}$ as
\begin{equation}
  P_{\mathrm{ann,D}} =
  \int_0^{\infty}\mathrm{d}t_d\,
  \Gamma_{\mathrm{decay}}e^{-\gamma_{\mathrm{decay}}t_d}
  \left(
    1-\exp\left[
      -\int_{t_d}^{\infty}\mathrm{d}t\,
      \Gamma_{\mathrm{ann,D}}(t)
    \right]
  \right).
\end{equation}
In this case, because $\int_{t_d}^{\infty}\mathrm{d}t\,\Gamma_{\mathrm{ann,D}}(t)$ is always much smaller than unity,
this equation can be approximated as
\begin{equation}
  P_{\mathrm{ann,D}} =
  \int_0^{\infty}\mathrm{d}t_d\,
  \Gamma_{\mathrm{decay}}e^{-\gamma_{\mathrm{decay}}t_d}
  \int_{t_d}^{\infty}\mathrm{d}t\,
  \Gamma_{\mathrm{ann,D}}(t).
\end{equation}
By assuming $a \propto t^s$, this can be rewritten as
\begin{equation}
  P_{\mathrm{ann,D}} =
  s^3\frac{\tilde{\Gamma}_{\mathrm{ann,D}}(t_1)}{\Gamma_{\mathrm{decay}}}
  \int_0^{\infty}\mathrm{d}x_d\, e^{-sx_d} x_d^s
  \int_{x_d}^{\infty}\mathrm{d}x\, x^{-5s}
  \int_0^x\mathrm{d}\tilde{x}\, e^{-s\tilde{x}} \tilde{x}^s.
  \label{eq:prob ann D}
\end{equation}

Next we calculate $\tilde{\Gamma}_{\mathrm{ann,D}}(t_1)$.
Assuming an instantaneous decay and that the number density of the high energy neutrinos from the $\nu_e + \pi^0$ channel is $B_{\nu_e \pi^0}n_{\nu_s}(t_1)/2$,
where the factor $1/2$ comes from the distinction of neutrinos and anti-neutrinos.
The cross section is
\begin{equation}
  \langle \sigma \rangle =
  \frac{8G_F^2 E_{\nu}^2 }{15\pi}(C_V^2 + C_A^2)~.
\end{equation}
Therefore
\begin{equation}
  \tilde{\Gamma}_{\mathrm{ann,D}}(t_1)  =
  \frac{ 4G_F^2 E_{\nu}^2 }{ 15\pi }B_{\nu_e\pi^0}(C_V^2 + C_A^2)n_{\nu_s}(t_1)~,
  \label{eq:ann rate D}
\end{equation}
where we used (\ref{eq:decay temperature of sterile neutrino}).

By using (\ref{eq:prob ann D}) and (\ref{eq:ann rate D}) and focusing on the $\nu_e + \pi^0$ channel, we can calculate $P_{\mathrm{ann,D}}$ and the contribution of this process to the correction to $\overline{f}$, $\Delta \overline{f}_{\mathrm{ann,D,\nu_e\pi^0}}$,
\begin{equation}
  \Delta \overline{f}_{\mathrm{ann,D},\nu_e\pi^0} =
  \left(\frac{1}{2} - \frac{m_{\pi^0}^2}{2m_s^2}\right)
  P_{\mathrm{ann,D}} B_{\nu_e\pi^0}.
\end{equation}

In Fig~\ref{fig:f correction}, we used $s = 2/3$, which gives larger corrections to $\overline{f}$ than $s = 1/2$.
We can check the contribution to the corrections to $\overline{f}$ from the other decay processes are less than $\Delta \overline{f}_{\mathrm{ann},\nu_e\pi^0}$ by similar calculations.
In this work, we assume that the correction to $\overline{f}$ can be regarded negligible in the region satisfying $\Delta \overline{f}_{\mathrm{ann},\nu_e\pi^0}<0.01$.

\section{BBN results in more detail}
\label{sec:BBN result appendix}
\begin{figure}[tb]
  \begin{center}
   	 \includegraphics[clip,width=0.44\textwidth]{./pics/BBNresults/000chi.png}
 	 \includegraphics[clip,width=0.44\textwidth]{./pics/BBNresults/000chicomb.png}
    \includegraphics[clip,width=0.44\textwidth]{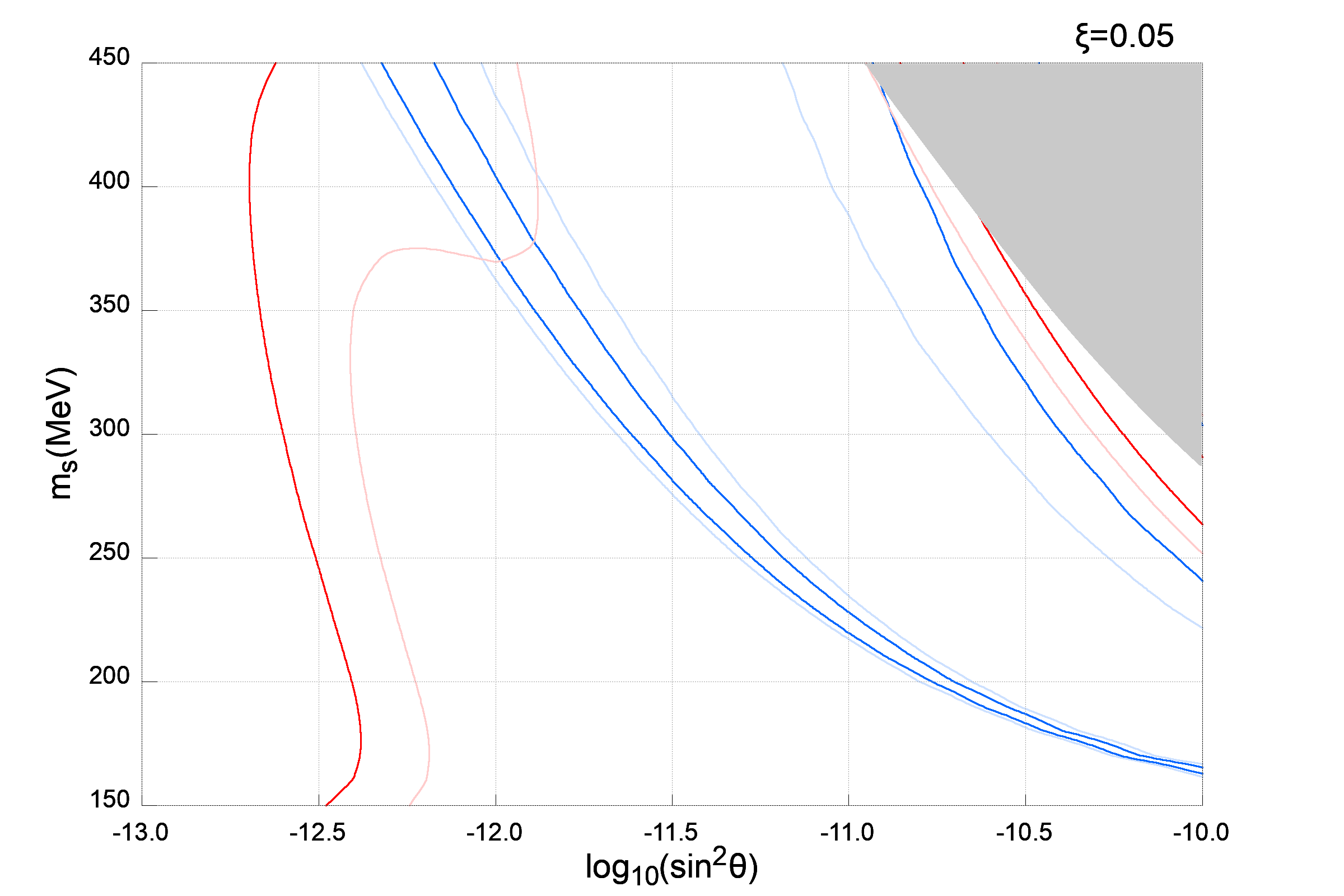}
    \includegraphics[clip,width=0.44\textwidth]{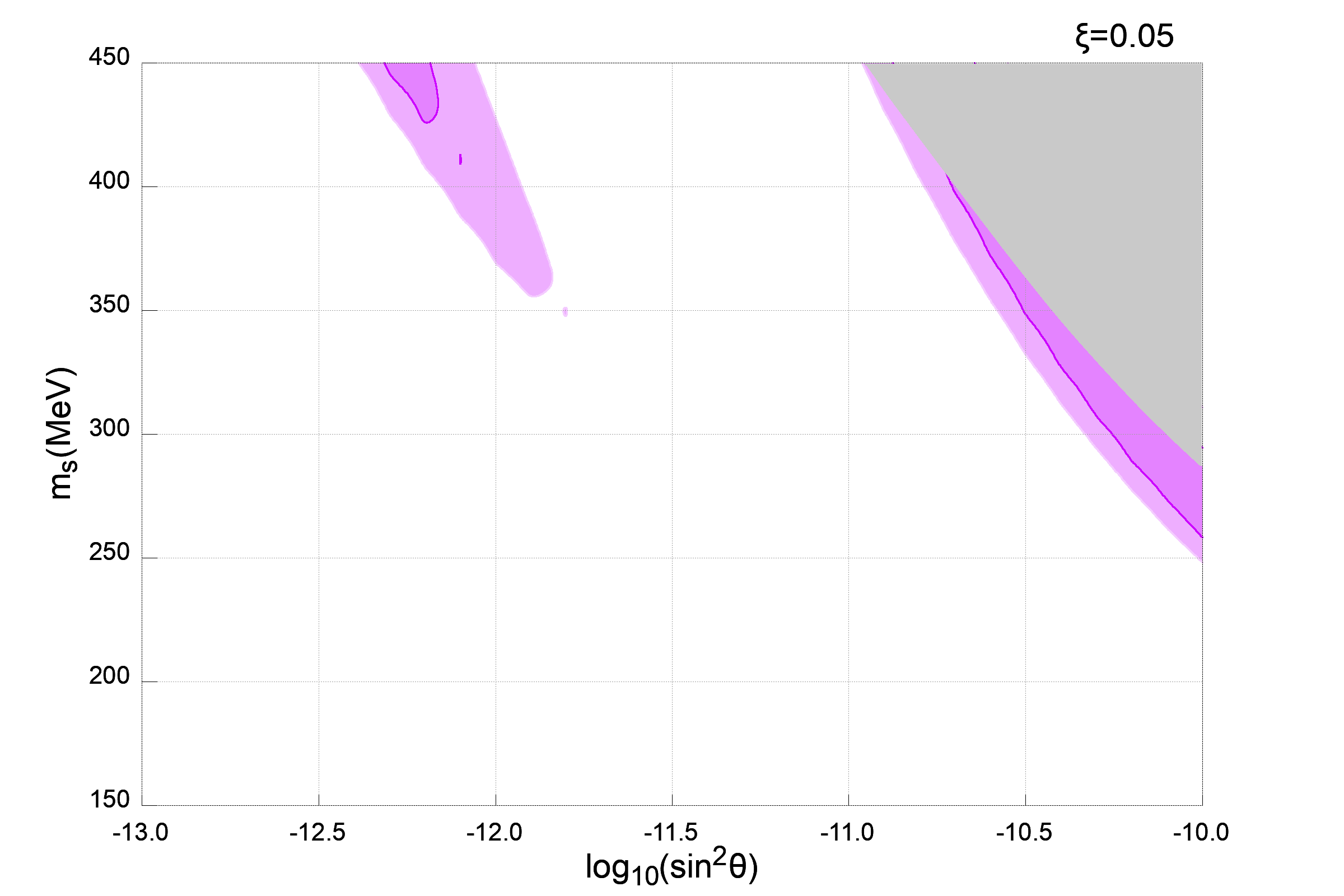}
    \includegraphics[clip,width=0.44\textwidth]{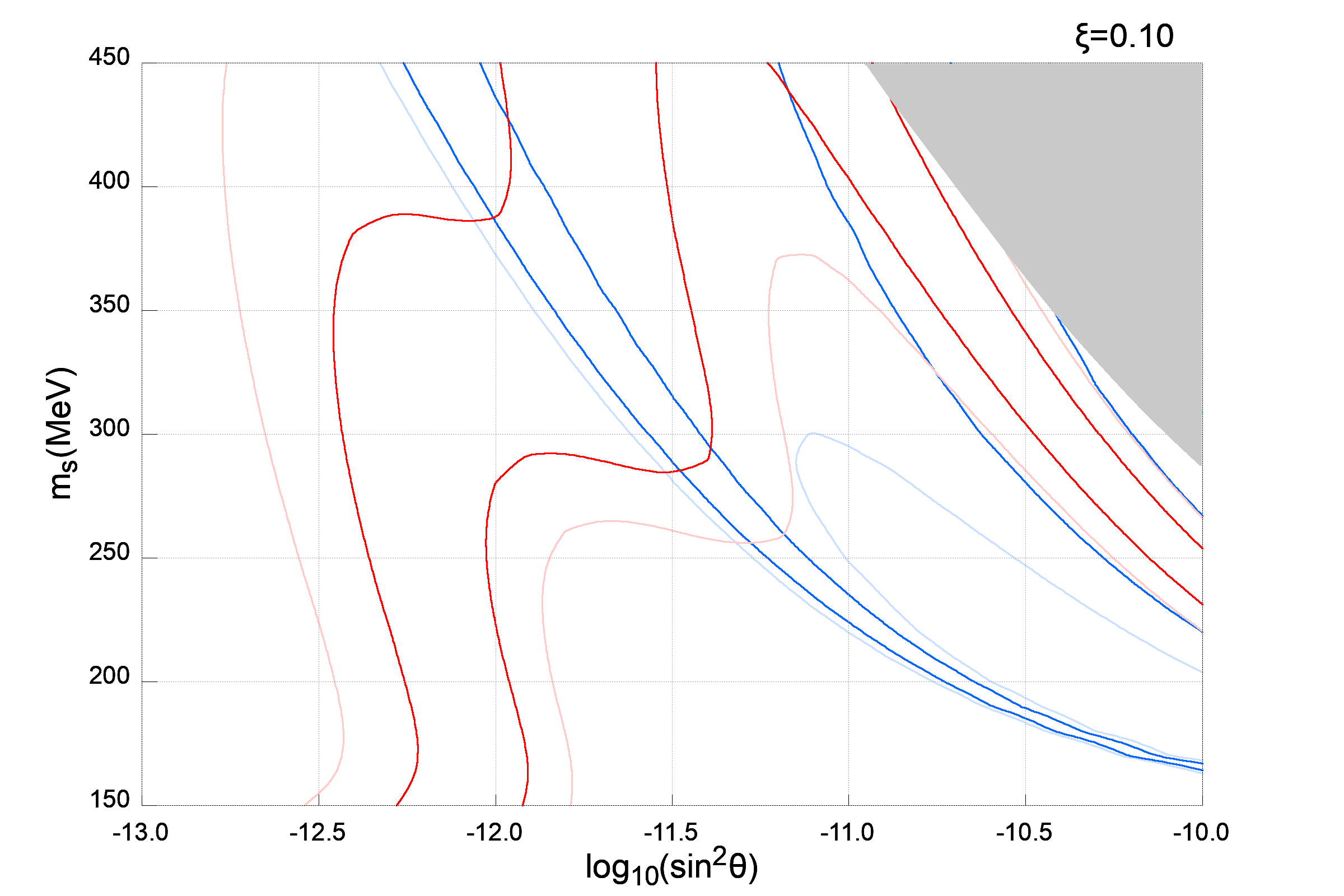}
    \includegraphics[clip,width=0.44\textwidth]{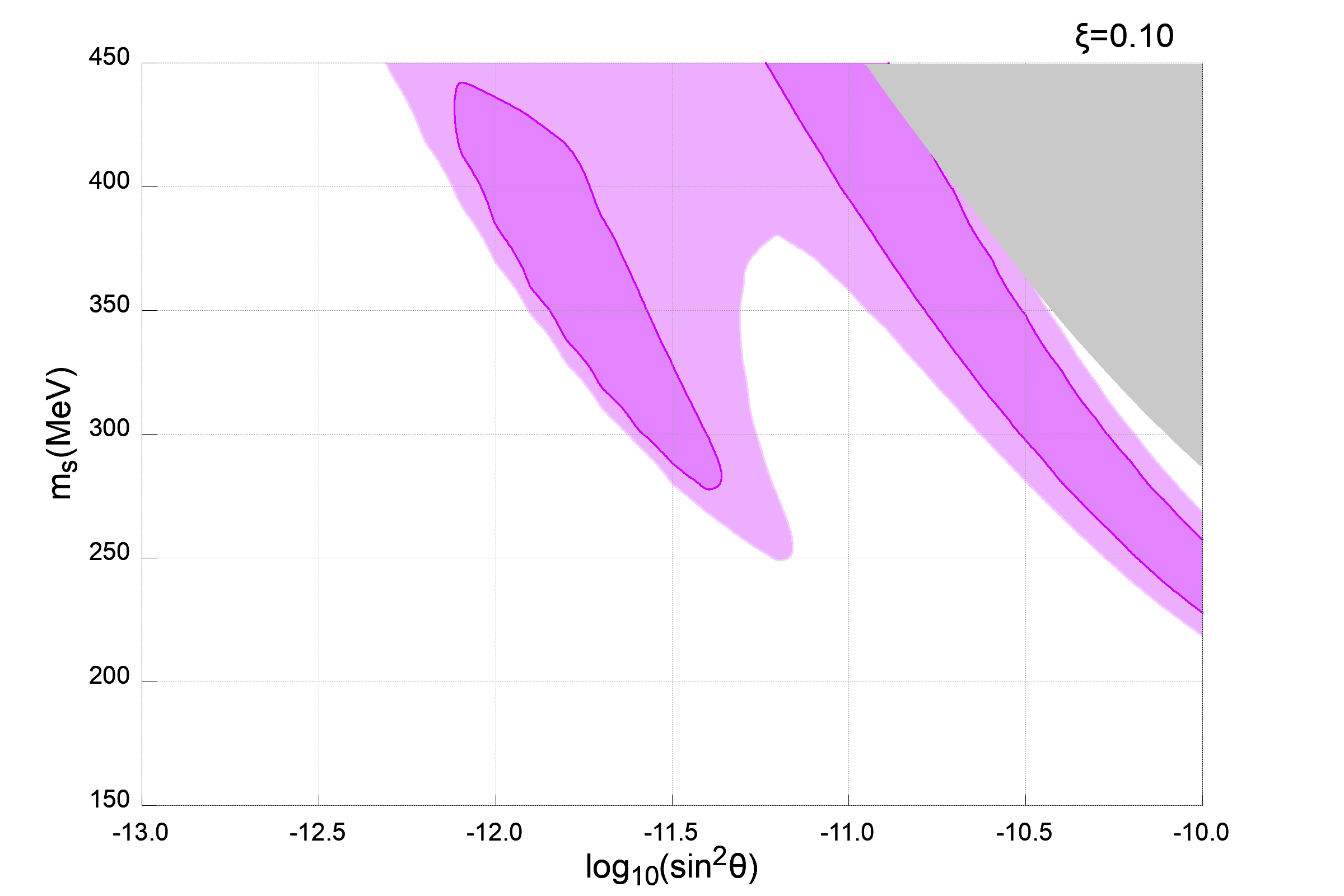}
    \includegraphics[clip,width=0.44\textwidth]{./pics/BBNresults/015chi.png}
    \includegraphics[clip,width=0.44\textwidth]{./pics/BBNresults/015chicomb.png}

     \caption{The result of the numerical calculation of BBN with the sterile neutrinos and lepton asymmetries from $\xi=0, 0.05, 0.1$ and  $0.15$.}
      \label{fig:BBNresult app1}
  \end{center}
\end{figure}

\begin{figure}[tb]
  \begin{center}
    \includegraphics[clip,width=0.44\textwidth]{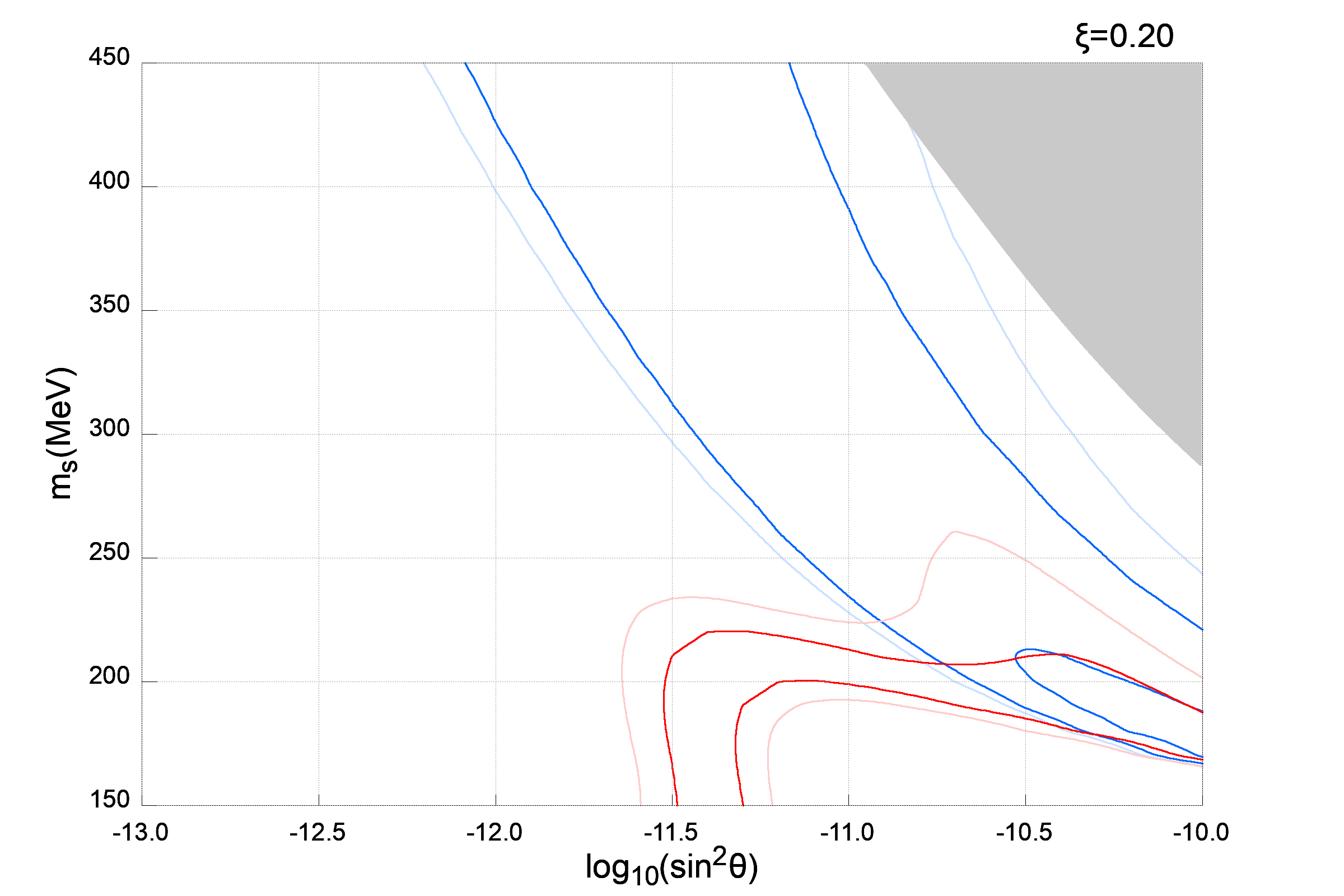}
    \includegraphics[clip,width=0.44\textwidth]{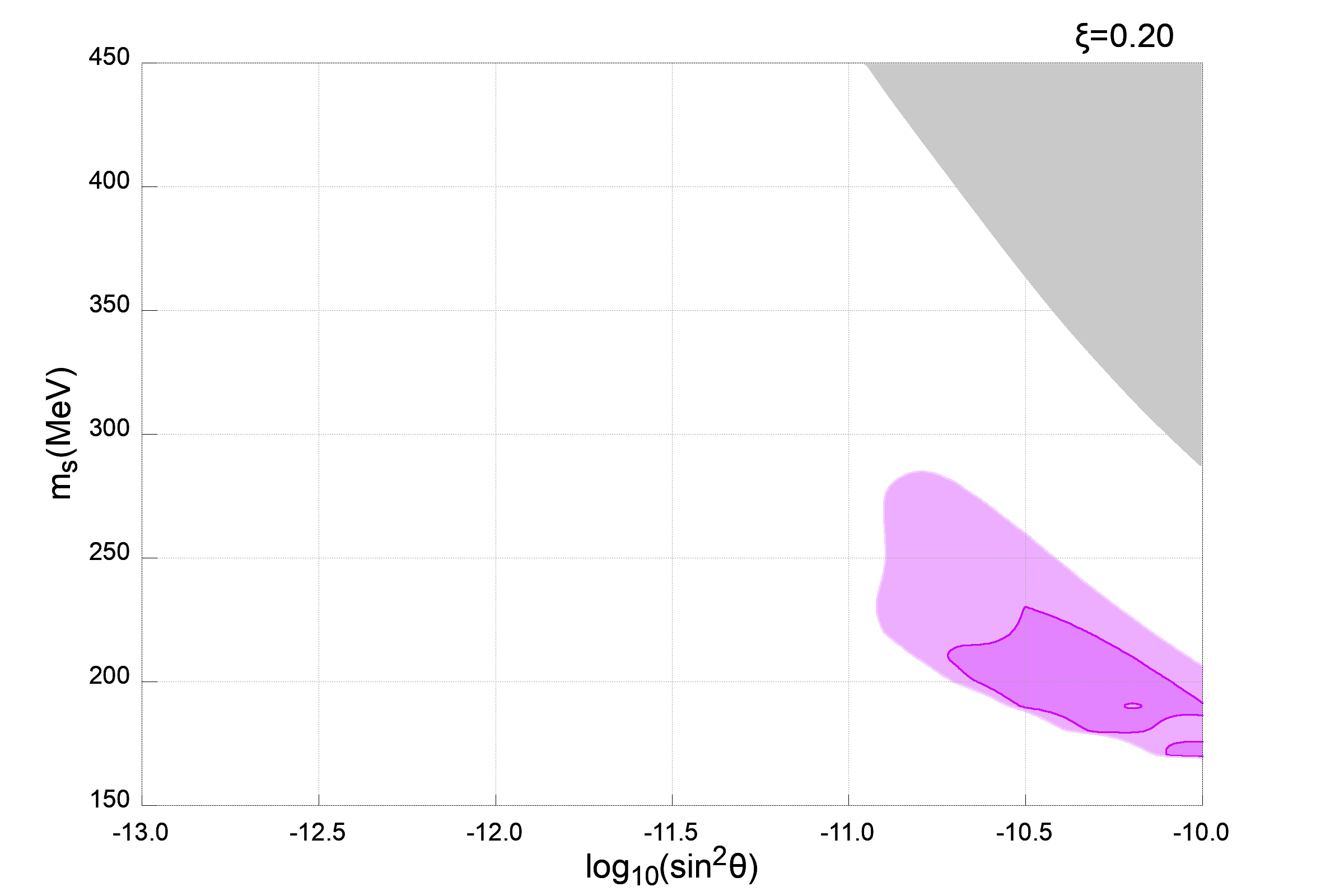}
    \includegraphics[clip,width=0.44\textwidth]{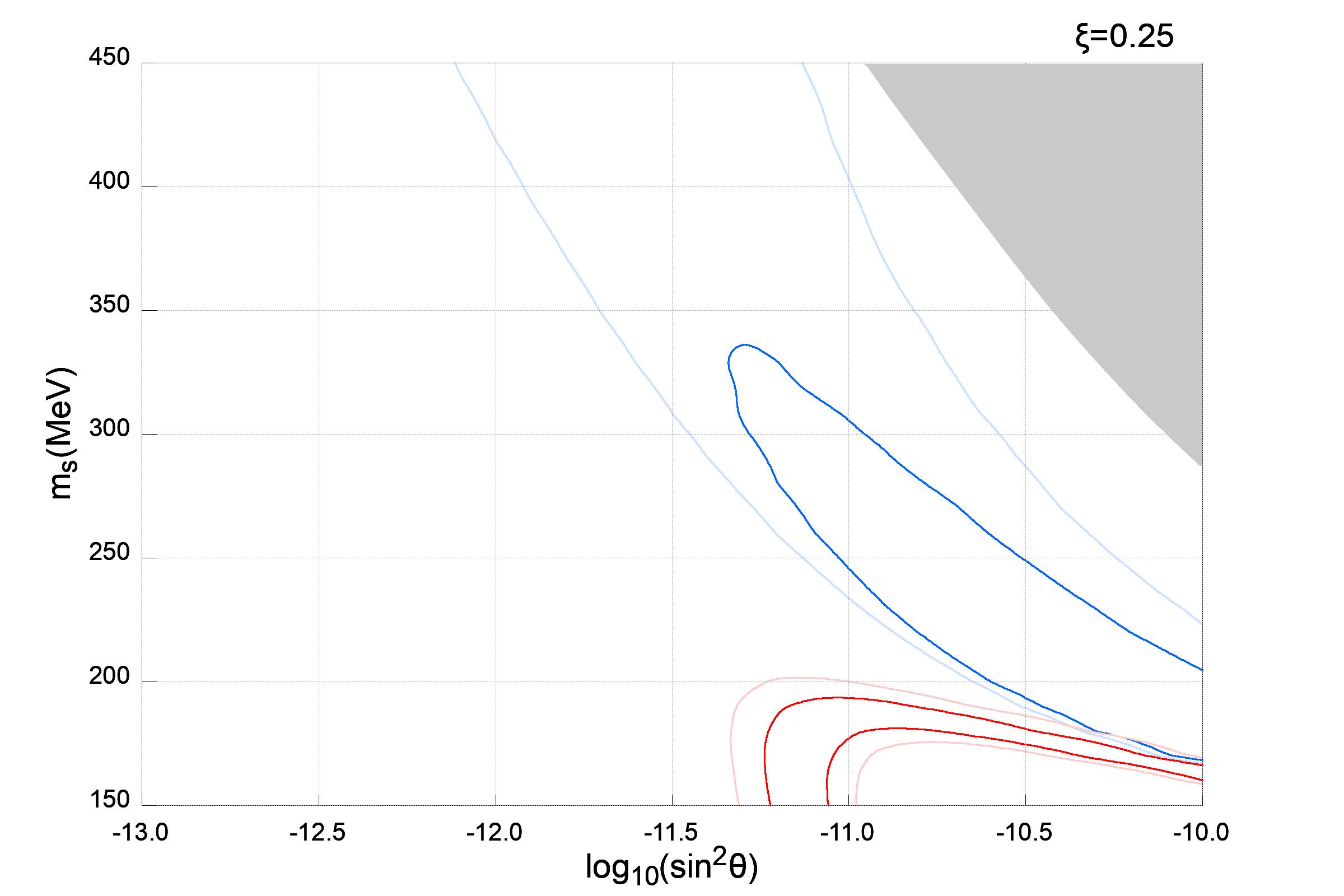}
    \includegraphics[clip,width=0.44\textwidth]{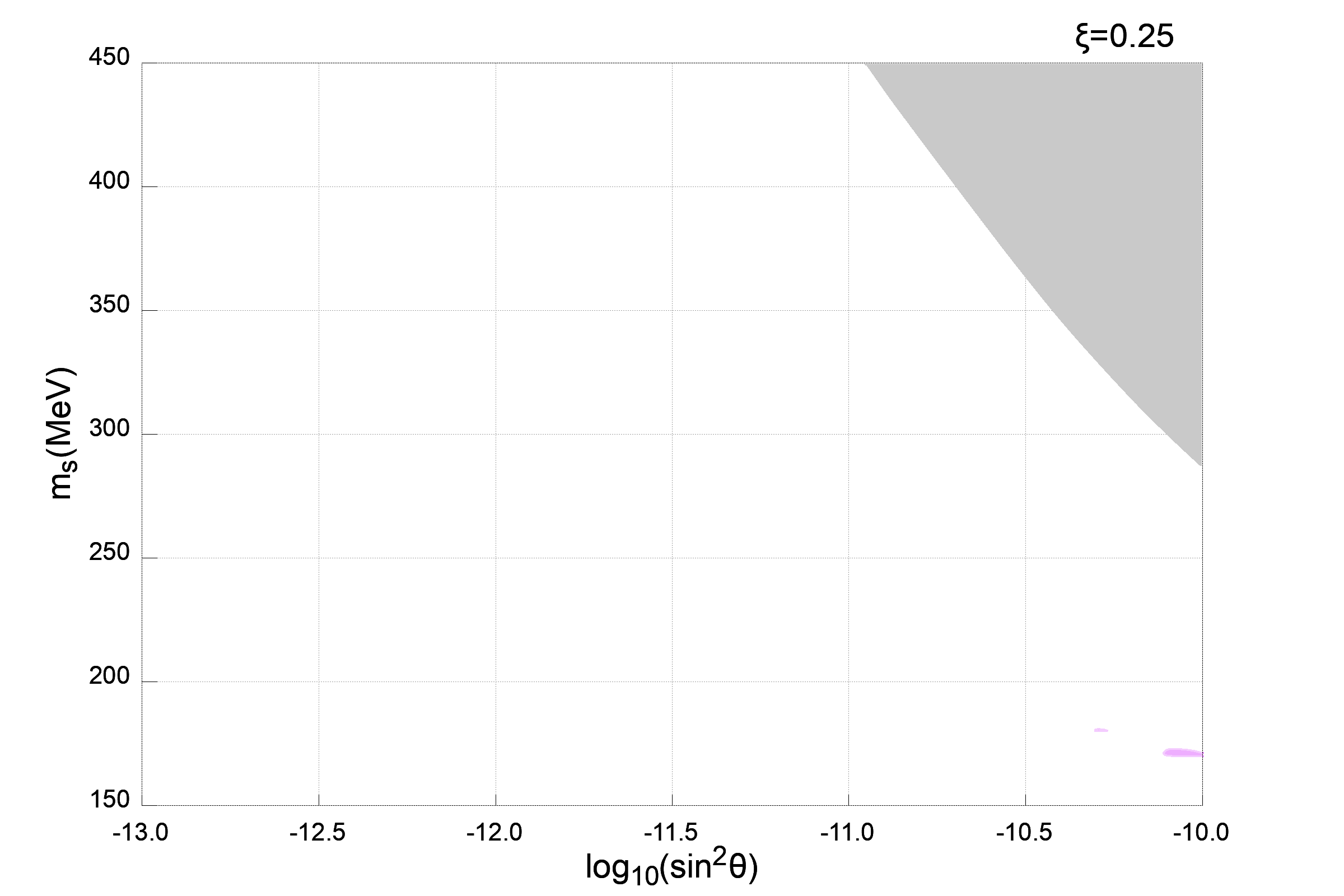}
    \caption{Same as Fig.~\ref{fig:BBNresult app1} except for  $\xi=0.2$ and $0.25$.}
    \label{fig:BBNresult app2}
  \end{center}
\end{figure}

Here we show the results of the numerical BBN calculation with lepton asymmetries in more detail and make some comments on the choice of the observational constraints. In
Fig.~\ref{fig:BBNresult app1} and \ref{fig:BBNresult app2}, we can find the regions that are consistent with the BBN constraints for $0.05 \lesssim \xi \lesssim 0.2$.

\begin{figure}[tbp]
 \begin{subfigure}[]{0.99\linewidth}
  \centering
  \includegraphics[clip,width=0.44\textwidth]{./pics/BBNresults/015chi.png}
  \includegraphics[clip,width=0.44\textwidth]{./pics/BBNresults/015chicomb.png}
  \caption{Cooke \textit{et al.}~\cite{Cooke:2017cwo} for D/H and Aver \textit{et al.}~\cite{Aver:2015iza} for $Y_P$}
 \end{subfigure}

 \vspace{5mm}

 \begin{subfigure}[]{0.99\linewidth}
  \centering
  \includegraphics[clip,width=0.44\textwidth]{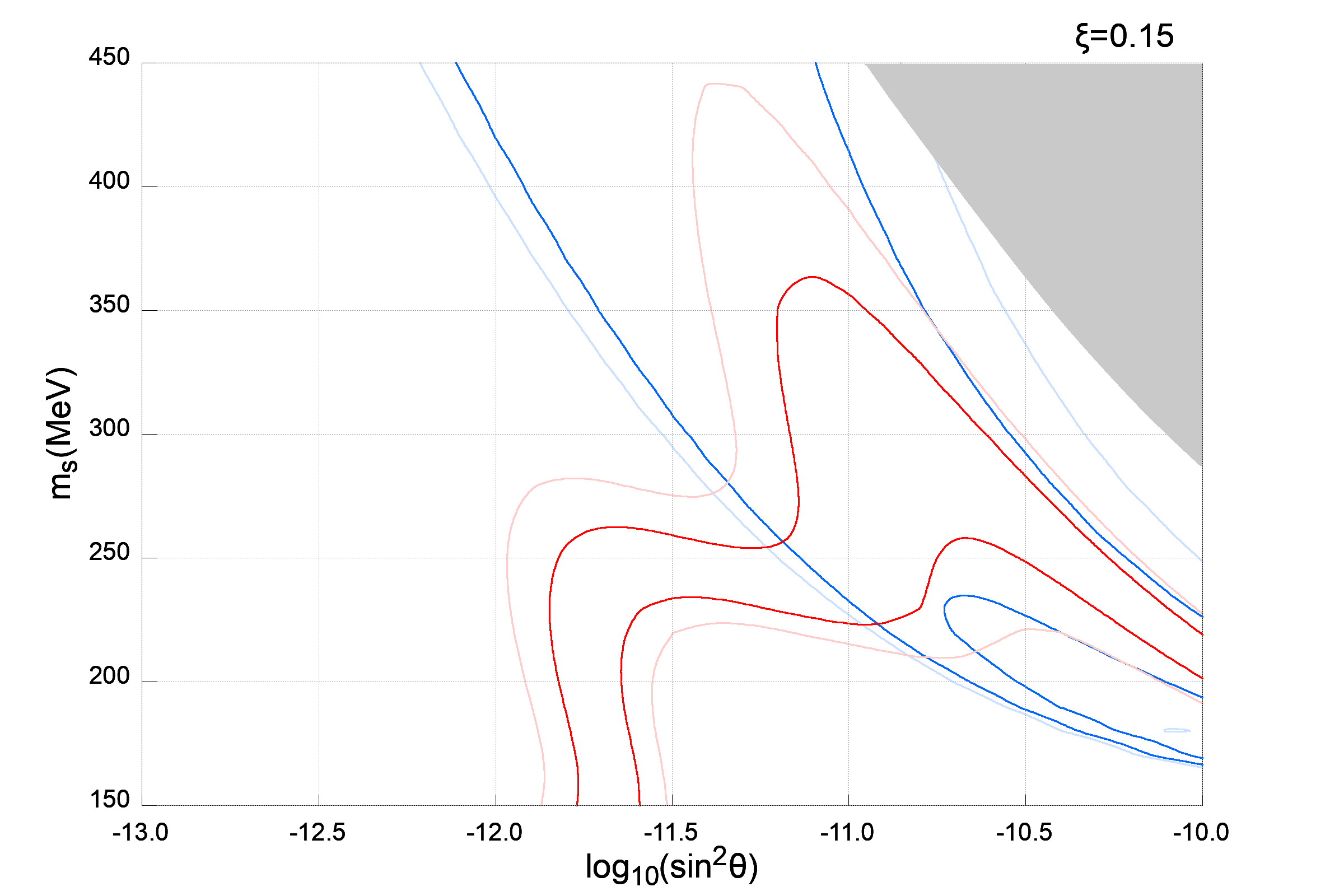}
  \includegraphics[clip,width=0.44\textwidth]{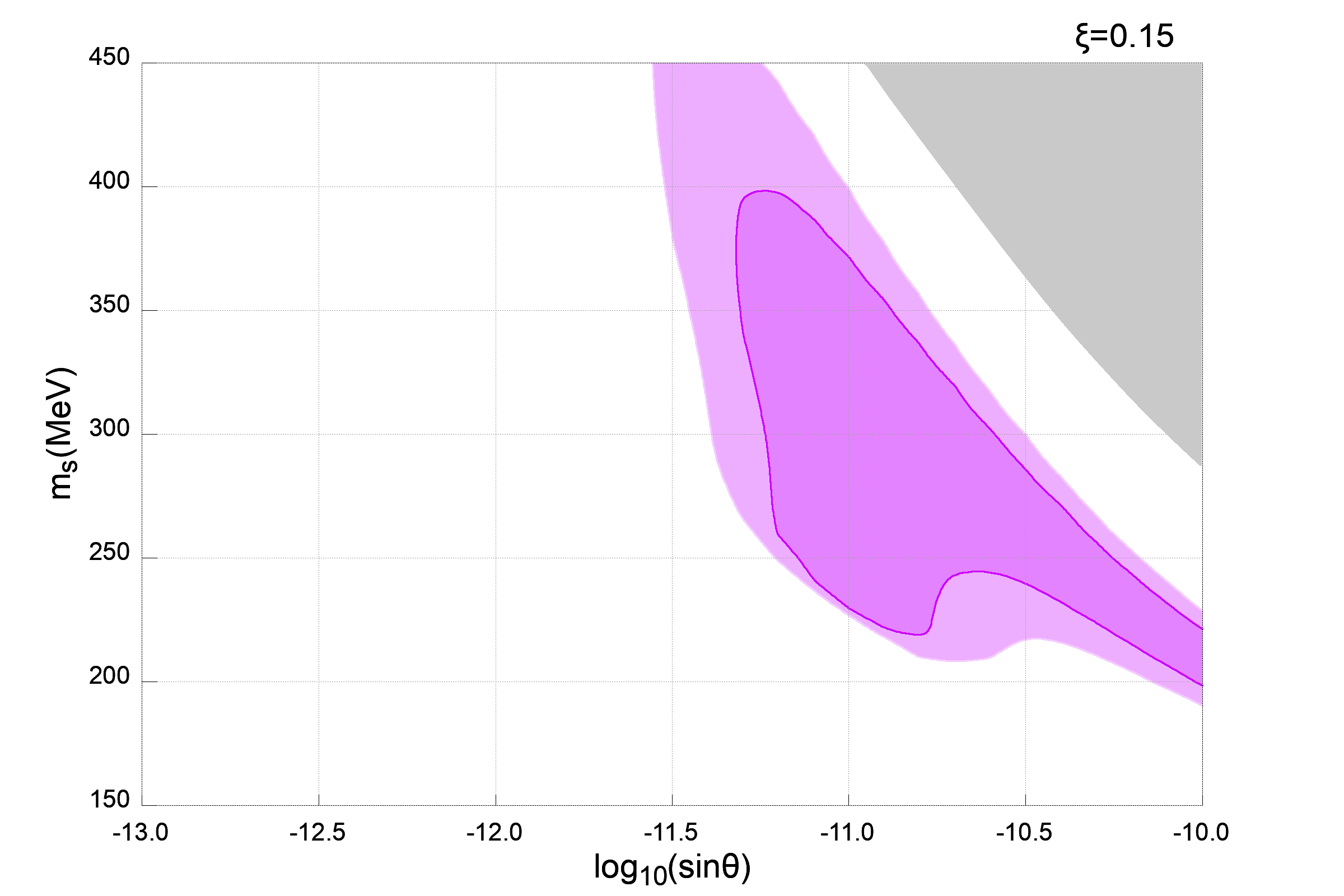}
  \caption{Particle Data Group~\cite{Tanabashi:2018oca} for D/H and $Y_P$}
 \end{subfigure}

 \vspace{5mm}

 \begin{subfigure}[]{0.99\linewidth}
  \centering
  \includegraphics[clip,width=0.44\textwidth]{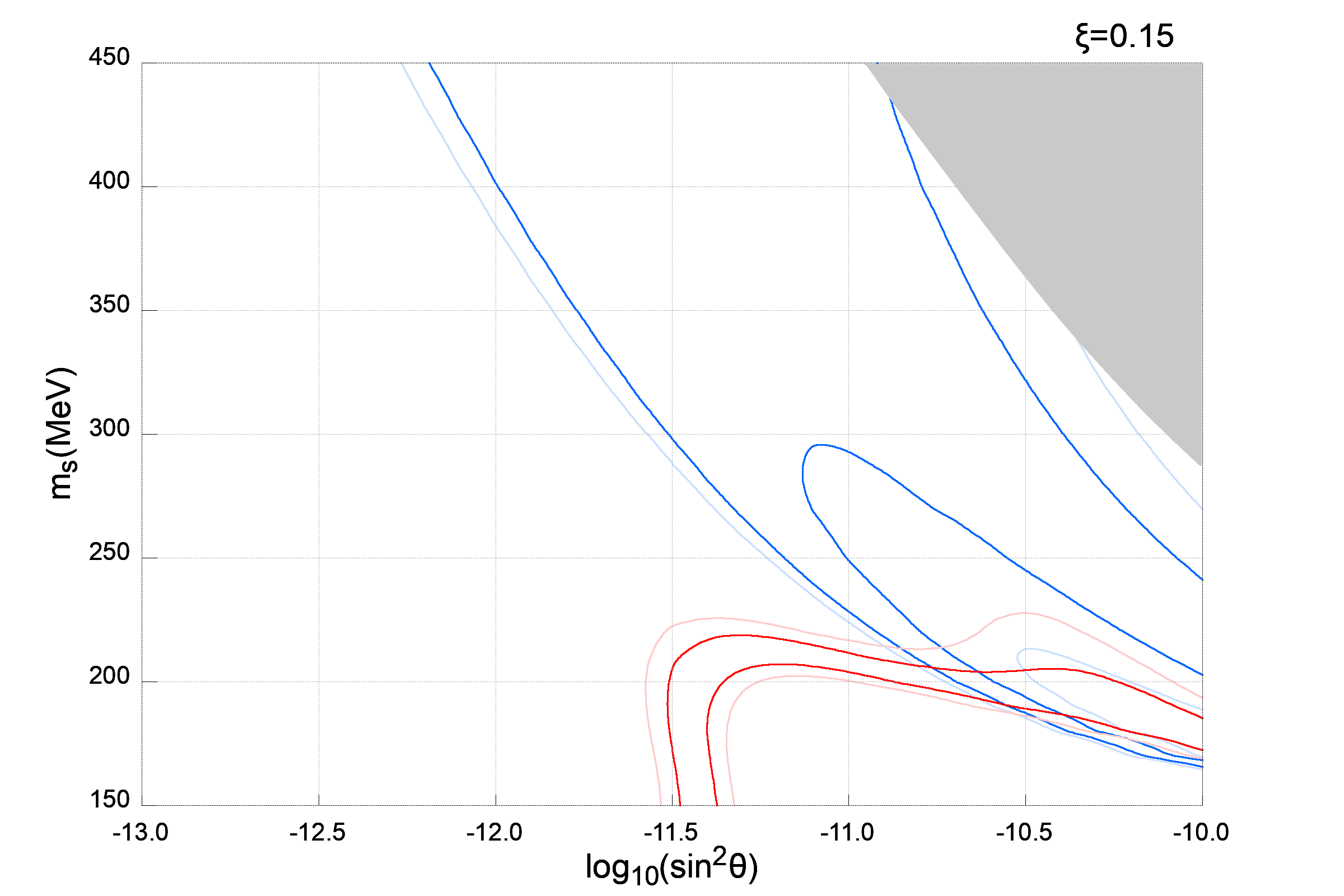}
  \includegraphics[clip,width=0.44\textwidth]{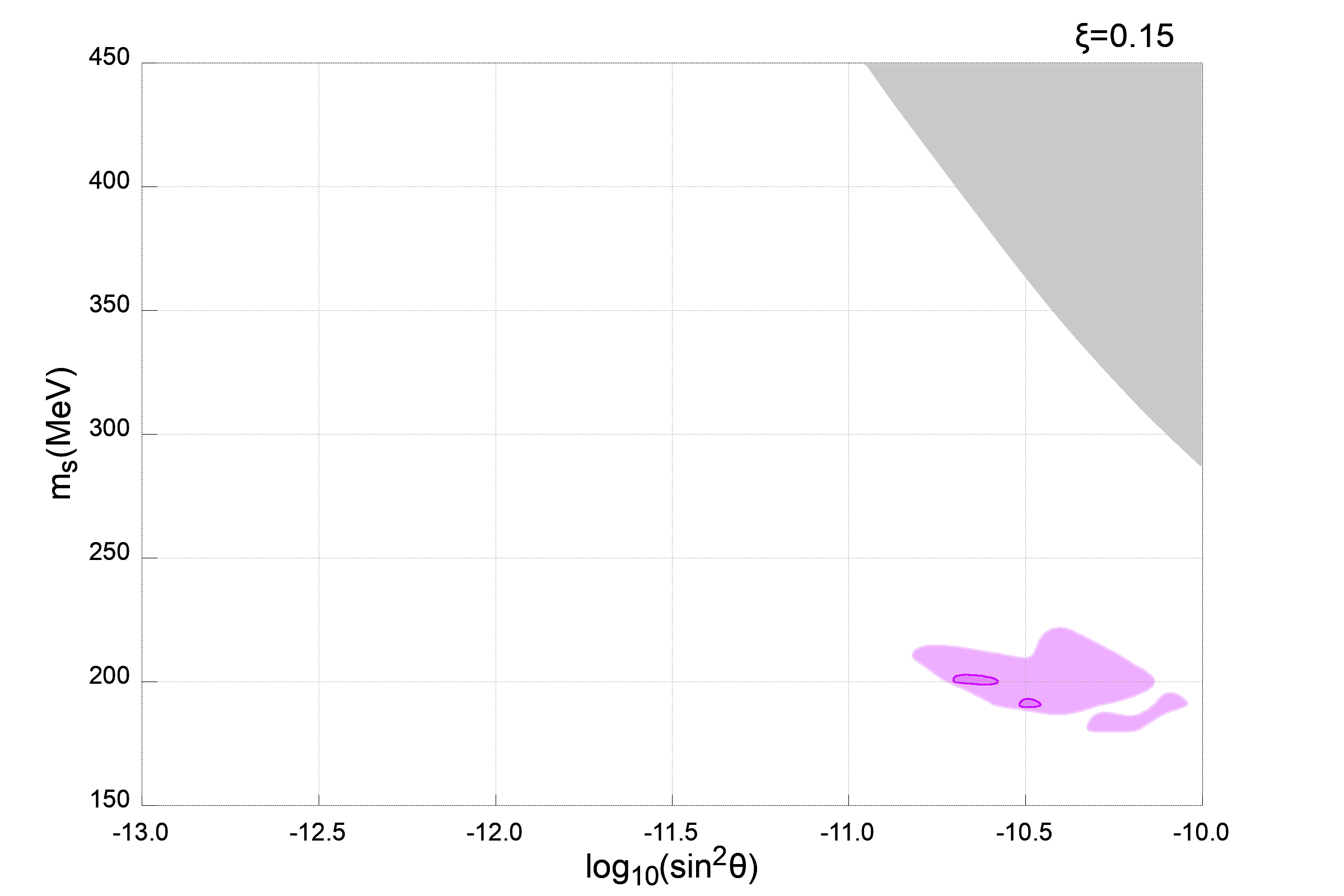}
  \caption{Cooke \textit{et al.}~\cite{Cooke:2017cwo} for D/H and Izotov \textit{et al.}~\cite{10.1093/mnras/stu1771} for $Y_P$}
 \end{subfigure}

 \caption{The comparison of the BBN results with different observational estimates.
 }
 \label{fig:BBN results with different obs}
\end{figure}

In  Sec.~\ref{sec:sterile BBN w/o lepton asymmetry} and~\ref{sec:sterile BBN w/ lepton asymmetry}, we adopt  Cooke \textit{et al.}~\cite{Cooke:2017cwo} for D/H and  Aver, Olive and Skillman~\cite{Aver:2015iza} for $Y_p$.
However we can consider several choices for the observational constraints.
In Fig.~\ref{fig:BBN results with different obs}, we compare the results when we adopt different observational constraints.
As can be seen from Fig.~\ref{fig:BBN results with different obs},
the existence of the consistent regions does not change.
However the constraints are more severe when we adopt Izotov \textit{et al.}~\cite{10.1093/mnras/stu1771} for the $Y_P$ limits.

\section{Generation of a large lepton asymmetry}
\label{sec:lepton_asymmetry}

In this section we present a model for generating a large lepton asymmetry.
The model is based on the Affleck-Dine mechanism in the minimal supersymmetric standard model (MSSM)~\cite{Affleck1985,Dine1996}.
In the MSSM there exist many flat directions in the scalar potential of squark, slepton and Higgs fields~\cite{Gherghetta:1995dv}.
Some flat directions have large field values during inflation.
Such flat directions start to oscillate after inflation and produce a baryon (lepton) asymmetry if they have a baryon (lepton) number.
This is called Affleck-Dine mechanism for baryogenesis or leptogenesis (for a review, see Ref.~\cite{Dine:2003ax}).
Here we focus on the Affleck-Dine leptogenesis in gauge-mediated SUSY breaking.

Suppose some flat direction (called AD field $\phi$) with lepton number has a large field value during inflation.
The potential of the AD field for $|\phi| \gg M_s$ is written as
\begin{align}
    V(\phi) =&~ V_\text{gauge} + V_\text{grav} + V_A \nonumber \\
    = &~ M_F^4 \left[\log\left(\frac{|\phi|^2}{M_s^2}\right)\right]^2
    + m_{3/2}^2 |\phi|^2 \left(1+ K \log\frac{|\phi|^2}{M^2}\right)
    + V_A~,
\end{align}
where $M_s$ is the messenger scale, $M_F$ is the SUSY breaking scale, $m_{3/2}$ is the gravitino mass, $K$ is the coefficient of the one-loop correction and $M$ is the renormalization scale.
Here $V_A$ is the A-term which generates the lepton asymmetry.
$V_\text{gauge}$ denotes the potential coming from the gauge-mediated SUSY breaking~\cite{deGouvea:1997afu} and $V_\text{grav}$ is due to the gravity mediation.
The potential is dominated by $V_\text{grav}$ for $|\phi| \gtrsim \varphi_\text{eq} \simeq M_F^2/m_{3/2}$.
We assume that the field value $\varphi =|\phi|$ during inflation is much larger than $\varphi_\text{eq}$.
When the Hubble parameter $H$ becomes equal to the effective mass, i.e. $H\simeq m_{3/2}$, the AD field starts to oscillate.
At the same time the AD field is kicked to the phase direction due to the A-term $V_A$, which leads to the generation of a lepton asymmetry.
The produced lepton asymmetry is given by
\begin{equation}
    \eta_L = \frac{n_L}{s} \simeq \varepsilon\frac{m_{3/2}\varphi_\text{osc}^2}{4m_{3/2}^2 M_p^2/T_R}
    = \varepsilon \frac{T_R}{4m_{3/2}}\left(\frac{\varphi_\text{osc}}{M_p}\right)^2,
    \label{eq:AD_lepton_asym}
\end{equation}
where $\varphi_\text{osc} > \varphi_\text{eq}$ is the field value at the start of oscillation, $T_R$ is the reheating temperature after inflation and $\varepsilon (\le 1)$ is the parameter which represents the efficiency of the A-term.
Here we assumed that the oscillation starts before reheating and used the relation between the entropy $s$ and total radiation densities $\rho_R$ given by $s = 4\rho_R/3T$.
Thus, the AD mechanism can generate a large lepton asymmetry.
In the following we take $\varepsilon =1$.

Another important consequence of the AD mechanism is the  formation of Q-balls~\cite{Coleman:1985ki,Kusenko:1997zq,Dvali:1997qv,Kusenko:1997si,Enqvist:1997si,Kasuya:1999wu}.
During oscillation the AD field fragments into spherical lumps through spatial instabilities of the field and almost all lepton number is confined within Q-balls.
When the potential is dominated by $V_\text{grav}$ the Q-ball formation depends on whether $K$ is positive or negative.
Since the instability of the AD field develops only for $K <0$, Q-balls are (not) formed for $K<0$ ($K>0$)~\cite{Kasuya:2000sc}.
On the other hand, when the potential is dominated by $V_\text{gauge}$, Q-balls are always produced.
Therefore, for $K > 0$ the Q-ball formation takes place when the AD field decreases to $\varphi_\text{eq}$.
The Q-ball in this case is called delayed~\cite{Kasuya:2001hg}.
In the following we show that the delayed Q-balls in the AD mechanism produce a large lepton asymmetry without significant entropy production.\footnote{
Another scenario with large entropy production was proposed in Ref.~\cite{Kawasaki:2002hq}.}

The Q-balls with large lepton number decay by emitting neutrinos with decay rate~\cite{Cohen:1986ct,Kawasaki:2012gk},
\begin{equation}
    \Gamma_Q \simeq N_\ell \frac{1}{Q} \frac{\omega_Q^3}{12\pi^2}
    4\pi R_Q^2,
\end{equation}
where $Q$ is the Q-ball charge ($=$ lepton number), $\omega_Q$ is the energy per charge, $R_Q$ is the Q-ball radius and $N_\ell$ is the number of decay channels.
$\omega_Q$, $R_Q$ and $Q$ are given by~\cite{Dvali:1997qv,Kusenko:1997si}
\begin{align}
    \omega_Q & \simeq M_F Q^{-1/4} ,\\
    R_Q & \simeq M_F^{-1} Q^{1/4} ,\\
    Q & \simeq 6\times 10^{-4} \left(\frac{\varphi_\text{eq}}{M_F}\right)^4 .
\end{align}
The cosmic temperature $T_\text{QD}$ when the Q-balls decay is estimated as
\begin{align}
    T_\text{QD} & \simeq \left(\frac{90}{\pi^2 g_*(T_\text{QD})}\right)^{1/4} \sqrt{M_{p}\Gamma_{Q}}\nonumber \\
    & \simeq 0.018 \text{GeV}\, \left(\frac{g_*}{60.75}\right)^{-1/4} \left(\frac{N_\ell}{3}\right)^{1/2}
    \left(\frac{M_F}{10^5\text{GeV}}\right)^{-2}
    \left(\frac{m_{3/2}}{0.1\text{GeV}}\right)^{5/2} ,
\end{align}
where $g_*$ is the number of degrees of freedom at $T_\text{QD}$.
The Q-ball energy density at $T_\text{QD}$ is written as
\begin{equation}
    \left. \frac{\rho_Q}{\rho_R} \right|_{T_\text{QD}}
    \simeq \frac{\omega_Q n_L}{(3/4)T_\text{QD} s(T_\text{QD})}
    \simeq  8.5 \frac{m_{3/2}}{T_\text{QD}}\eta_L .
\end{equation}
From Eq.~(\ref{eq:AD_lepton_asym}) the lepton asymmetry $\eta_L$ is written as
\begin{equation}
    \eta_L \simeq 0.42 \left(\frac{T_R}{10^4\text{GeV}}\right)
    \left(\frac{M_F}{10^5\text{GeV}}\right)^4
    \left(\frac{m_{3/2}}{0.1\text{GeV}}\right)^{-3}
    \left(\frac{\varphi_\text{osc}}{10^5 \varphi_\text{eq}}\right)^2 .
\end{equation}
For example, for $M_F=0.9\times 10^5$~GeV, $T_R =1.5\times 10^5$~GeV, $m_{3/2}=0.6$~GeV and $\varphi_\text{osc}=10^5\varphi_\text{eq}\simeq  1.4\times 10^{15}$~GeV, we obtain $\rho_Q/\rho_R \simeq 0.05$ and $\eta_L \simeq 0.02$.
Therefore, a large lepton asymmetry is produced with negligible entropy production.

Lepton asymmetry is also produced from the Q-balls through evaporation~\cite{Laine:1998rg,Banerjee:2000mb}.
The evaporated charge $\Delta Q$ is written as~\cite{Kasuya:2014ofa}
\begin{equation}
    \frac{\Delta Q}{Q} \sim 10^{-3}\left(\frac{M_s}{10^4\text{GeV}}\right)^{-2/3}
    \left(\frac{M_F}{10^5\text{GeV}}\right)^{-4}
    \left(\frac{m_{3/2}}{0.1\text{GeV}}\right)^{11/3},
    \label{eq:Qball_evaporation}
\end{equation}
where $M_s$ is the sparticle mass and the evaporation is most efficient around $T_* ( < T_R)$ given by\footnote{
The evaporated charge after the electroweak phase transition is suppressed by $(T_\text{EW}/T_*)^2$ compared with Eq.~(\ref{eq:Qball_evaporation})  where $T_\text{EW}$ is the temperature at the electroweak phase transition.}
\begin{equation}
    T_* \sim 10^3 \text{GeV}
    \left(\frac{M_s}{10^4\text{GeV}}\right)^{2/3}
    \left(\frac{m_{3/2}}{0.1\text{GeV}}\right)^{1/3}.
\end{equation}
(For $M_F = 0.9\times 10^5$~GeV and $m_{3/2} = 0.6$~GeV, $\Delta Q/Q \sim 0.1$.)
Thus, some fraction of the lepton charge of the Q-balls are emitted into the thermal plasma before the electroweak phase transition, which leads to a baryon asymmetry through the sphaleron effect.
The produced baryon asymmetry $\eta_B$  by the Q-ball evaporation is estimated as $\eta_B \sim (\Delta Q/Q)\eta_L$ which is too large for $\eta_L \sim 0.01$.

In order to avoid this problem, we need some mechanism to wash out the lepton asymmetry before the electroweak phase transition.
This can be realized by introducing another sterile neutrino $\nu_h$ with a mass of $m_{hs} \sim 200$~GeV.
Before the electroweak phase transition, sterile neutrinos interact with Higgs bosons and left-handed leptons through the Yukawa interaction and inverse decays of leptons and Higgs bosons into sterile neutrinos can wash out the lepton asymmetry contained in active neutrinos.
The rate of washout $\Gamma_{\mathrm{w.o.}}$ can be estimated as~\cite{Buchmuller:2004nz,Buchmuller:2005eh}
\begin{equation}
  \Gamma_{\mathrm{w.o.}} \sim \frac{\lambda ^2}{16\pi}m_{hs}\frac{K_1(m_{hs}/T)}{K_2(m_{hs}/T)}
  = \frac{m_{hs}^3}{16\pi v^2}\frac{K_1(m_{hs}/T)}{K_2(m_{hs}/T)} \sin^2 \theta_h,
\end{equation}
where $\lambda$ is the coupling constant of the Yukawa interaction,
$K_i$ is the modified Bessel function of order $i$,
$v \simeq 246$~GeV is the vacuum expectation value of the Higgs field after the electroweak phase transition,
and $\theta_{h}$ is the $\nu_h$-$\nu_{\alpha}$ mixing angle in the vacuum.
In order to wash out the lepton asymmetry from the Q-ball evaporation, the washout process should be efficient just before the electroweak phase transition.
Therefore, we require $\Gamma_{\mathrm{w.o.}} \gg H$ at $T\sim 200$~GeV and derive the condition $\sin ^2 \theta_h \gg 8 \times 10^{-14}$.
In other words, if $m_h \sim 200$~GeV and $\sin^2 \theta_h \gg 8 \times 10^{-14}$,
the lepton asymmetry from the Q-ball evaporation will be washed out to the extent consistent with the observed baryon asymmetry.
This parameter region is free from observational or cosmological constraints~\cite{Kusenko:2009up,Drewes:2015iva}.

\bibliography{SterileBBN}
\bibliographystyle{JHEP}

\end{document}